\documentclass[iop]{emulateapj}
\usepackage{color}
\usepackage{graphicx}
\usepackage[font=small,labelfont=bf]{caption}
\usepackage{subfigure}
\usepackage{epsfig}
\usepackage{enumerate}
\usepackage{amsmath}
\usepackage{booktabs}
\usepackage{mathtools}
\newcommand{\az}{$|z|$}
\newcommand{\s}{\hspace{1 mm}}
\begin{document}
\DeclareGraphicsExtensions{.ps,.pdf,.png,.jpg}
\title{Chemical Cartography with APOGEE: Large-scale Mean Metallicity Maps of the Milky Way}

\author{Michael R. Hayden\altaffilmark{1}, 
Jon A. Holtzman\altaffilmark{1}, 
Jo Bovy\altaffilmark{2,23},
Steven R. Majewski\altaffilmark{3}, 
Carlos Allende Prieto\altaffilmark{4},
Timothy C. Beers\altaffilmark{5},
Katia Cunha\altaffilmark{6,7},                    
Peter M. Frinchaboy\altaffilmark{8},
Ana E. Garc\'{\i}a P\'erez\altaffilmark{3},
L\'eo Girardi \altaffilmark{9,10}, 
Fred R. Hearty \altaffilmark{11},
Jennifer A. Johnson\altaffilmark{12}, 
Young Sun Lee\altaffilmark{1},
David Nidever\altaffilmark{13},
Ricardo P. Schiavon\altaffilmark{14},
Katharine J. Schlesinger\altaffilmark{15},
Donald P. Schneider\altaffilmark{11,16},
Mathias Schultheis\altaffilmark{17},
Matthew Shetrone\altaffilmark{18},
Verne V. Smith\altaffilmark{19,7},
Gail Zasowski\altaffilmark{20},
Dmitry Bizyaev\altaffilmark{21},  
Diane Feuillet\altaffilmark{1},
Sten Hasselquist\altaffilmark{1},
Karen  Kinemuchi\altaffilmark{21},
Elena  Malanushenko\altaffilmark{21}, 
Viktor Malanushenko\altaffilmark{21},  
Robert O'Connell\altaffilmark{3},
Kaike Pan\altaffilmark{21},      
Keivan Stassun\altaffilmark{22}     
}

\altaffiltext{1}{New Mexico State University, Las Cruces, NM 88003, USA (mrhayden, holtz, yslee, feuilldk, sten@nmsu.edu)}

\altaffiltext{2}{Institute for Advanced Study, Einstein Drive, Princeton, NJ 08540, USA (bovy@ias.edu)}

\altaffiltext{3}{Dept. of Astronomy, University of Virginia, 
Charlottesville, VA 22904-4325, USA (srm4n, aeg4x, rwo@virginia.edu)}

\altaffiltext{4}{Instituto de Astrof\'{\i}sica de Canarias, 38205 La Laguna, Tenerife, Spain;
Departamento de Astrof\'{\i}sica, Universidad de La Laguna,
38206 La Laguna, Tenerife, Spain (callende@iac.es)}

\altaffiltext{5}{Department of Physics \& Astronomy and JINA, Joint Institute for Nuclear Astrophysics, Michigan State University, E. Lansing, MI  48824, USA (beers@pa.msu.edu)}

\altaffiltext{6}{Observat\'orio Nacional, S\~ao Crist\'ov\~ao, Rio de Janeiro, Brazil (cunha@email.noao.edu)}

\altaffiltext{7}{Steward Observatory, University of Arizona, Tucson, AZ 85721, USA} 

\altaffiltext{8}{Texas Christian University, Fort Worth, TX 76129, USA (p.frinchaboy@tcu.edu)}

\altaffiltext{9}{Osservatorio Astronomico di Padova-- INAF, Vicolo dell'Osservatorio 5, I-35122 Padova, Italy (leo.girardi@oapd.inaf.it)}

\altaffiltext{10}{Laborat\'orio Interinstitucional de e-Astronomia - LIneA, Rua Gal. Jos\'e Cristino 77, Rio de Janeiro, RJ - 20921-400, Brazil}

\altaffiltext{11}{Department of Astronomy and Astrophysics, The Pennsylvania State University, University Park, PA 16802, USA (frh10,dps7@psu.edu)}

\altaffiltext{12}{Department of Astronomy, The Ohio State University, Columbus, OH 43210, USA (jaj@astronomy.ohio-state.edu)}

\altaffiltext{13}{Department of Astronomy, University of Michigan, Ann Arbor, MI 48109, USA (dnidever@umich.edu)}

\altaffiltext{14}{Astrophysics Research Institute, Liverpool John Moores University, Twelve Quays House, Egerton Wharf, Birkenhead CH41 ILD, United Kingdom (rpschiavon@gmail.com)}

\altaffiltext{15}{Research School of Astronomy \& Astrophysics, Australian National University, Mt. Stromlo Observatory, Cotter Rd, Weston, ACT 2611, Australia (katharine.schlesinger@anu.edu.au)}

\altaffiltext{16}{Institute for Gravitation and the Cosmos, The Pennsylvania State University, University Park, PA 16802, USA}

\altaffiltext{17}{Observatoire de la Cote d'Azur, Boulevard de l'Observatoire, B.P. 4229, F-06304 NICE Cedex 4, France (mathias.schultheis@oca.eu)}

\altaffiltext{18}{University of Texas at Austin, McDonald Observatory, Fort Davis, TX 79734, USA (shetrone@astro.as.utexas.edu)}

\altaffiltext{19}{National Optical Astronomy Observatories, Tucson, AZ 85719, USA (vsmith@email.noao.edu)}

\altaffiltext{20}{Department of Physics and Astronomy, Johns Hopkins University, Baltimore, MD 21218, USA (gail.zasowski@gmail.com)}

\altaffiltext{21}{Apache Point Observatory, P.O. Box 59, Sunspot, NM 88349-0059, USA (dmbiz, kinemuchi, elenam, viktorm, kpan@apo.nmsu.edu)}

\altaffiltext{22}{Department of Physics and Astronomy, Vanderbilt Universiy, 6301 Stevenson Center, VU Station B \#351807, Nashville, TN 37235, USA (keivan.stassun@vanderbilt.edu)}

\altaffiltext{23}{Hubble Fellow}


\begin{abstract}
We present Galactic mean metallicity maps derived from the first year of the
SDSS-III APOGEE experiment.  Mean abundances in different zones of
Galactocentric radius ($0 < R < 15$ kpc) at a range of heights
above the plane ($0<|z|<3$ kpc), are derived from a sample of nearly 20,000 stars with unprecedented coverage, including stars in the Galactic
mid-plane at large distances.
We also split the sample into subsamples of stars with low and
high-[$\alpha$/M] abundance ratios.  We assess possible biases 
in deriving the mean
abundances, and find they are likely to be small except in the inner
regions of the Galaxy. A negative radial gradient
exists over much of the Galaxy; however, the gradient appears to flatten
for $R<6$ kpc, in particular near the Galactic mid-plane and for 
low-[$\alpha$/M] stars. At $R>6$ kpc, the gradient flattens as one moves 
off of the plane, and is flatter at all heights for high-[$\alpha$/M] stars
than for low-[$\alpha$/M] stars. Alternatively, these gradients
can be described as vertical gradients that flatten at larger
Galactocentric radius; these vertical gradients are similar for both
low and high-[$\alpha$/M] populations. Stars with higher [$\alpha$/M] appear
to have a flatter radial gradient than stars with lower [$\alpha$/M]. This
could suggest that the metallicity gradient has grown steeper with time or, 
alternatively, that gradients are washed out over time by migration of stars.

\end{abstract}

\section{Introduction} 

The variation of stellar chemical abundances within galaxies contains information
about how galaxies are assembled, tracking processes such as gas
accretion, star formation timescales, and stellar migration. Within
galaxy disks, measurements of abundance gradients have the potential
to distinguish between different models of disk formation.  Many models
predict that galaxies form inside-out, with star formation proceeding more
rapidly, and perhaps starting earlier, in the inner regions of disks
(e.g., \citealt{Larson1976, Kobayashi2011, Bird2013}).  Models
of chemical and dynamical evolution are becoming increasingly sophisticated, and
it is now recognized that the structure of disks likely evolves with
time, and that migration of stars within disks may be important to understanding the observed chemical structure of the disk (e.g.,
\citealt{Sellwood2002, Schonrich2009a, Loebman2011, Bird2013,Kubryk2013}).

The Milky Way provides the opportunity to measure large numbers of stellar abundances on a
star-by-star basis, and chemical abundance gradients have been observed
in the Milky Way disk in both the radial and vertical directions.
Various tracers have been used including Cepheid variables (e.g., \citealt{Luck2011a,Lemasle2013a}),
planetary nebulae (e.g., \citealt{Henry2010,Stanghellini2010}),
HII regions (e.g., \citealt{Balser2011}), open
clusters (e.g., \citealt{Carrera2011,Frinchaboy2013}), B stars
(e.g., \citealt{Daflon2004,Daflon2009}), and surveys of main
sequence stars (e.g., \citealt{Cheng2012b} for the Sloan Extension
for Galactic Understanding and Exploration [SEGUE; \citealt{Yanny2009}] of the
Sloan Digital Sky Survey [SDSS; \citealt{York2000}]; 
the Geneva-Copenhagen Survey [\citealt{Nordstrom2004}]; \citealt{Boeche2013} for the Radial Velocity Experiment [RAVE; \citealt{Steinmetz2006}]). The measured
amplitude of the Milky Way radial gradient range from $-0.04$ dex kpc$^{-1}$ (OB Stars; \citealt{Daflon2009}) to $-0.099$ dex kpc$^{-1}$ (intermediate-age main sequence stars from GCS; \citealt{Nordstrom2004}). To the
extent to which different
tracers probe populations of different ages, comparison of measured metallicity
gradients have the potential to provide information on the evolution
of chemical gradients with time. However, 
there is still disagreement on whether the observed gradients steepen or flatten with time, with
some studies finding the radial gradient steepens with time \citep{Stanghellini2010}, and others 
reported that the gradient has flattened with time \citep{Maciel2009a}. Recent
models and simulations that include radial migration predict that observed gradients will be flatter for older populations,
as migration washes out the gradients of older populations (e.g., \citealt{Roskar2008,Loebman2011,Kubryk2013}). Direct 
comparison between different samples still remains problematic; 
a uniform sample spanning a range of ages is needed to help address
these on-going questions.

The behavior of the radial gradient with location has also been found to vary with different populations.  Open cluster measurements \citep{Yong2012,Frinchaboy2013} have found a flattening of the radial gradient at large Galactocentric radii, while observations with Cepheid variables \citep{Luck2011b,Lemasle2013a} find that the slope does not change in the outer disk.  In the inner Galaxy, \citet{Henry2010} find that the slope of the radial gradient is shallower than at the solar circle using planetary nebula, but Cepheid observations \citep{Luck2011b,Genovali2013} find that the gradient is steepest in the inner Galaxy. These previous studies have generally been limited in sample size to less than a few hundred objects, and often cover only a limited radial range within the Galaxy.

Vertical metallicity gradients have also been characterized near the solar circle
(e.g., \citealt{Hartkopf1982, Chen2003, Marsakov2006, Chen2011})
for both 
thin and thick-disk populations, although there
is a range in reported slopes. The slope of the 
vertical gradient has been measured to be $-0.22$ dex kpc$^{-1}$
\citep{Ak2007} for $|z|<3$ kpc in a sample of high-latitude G dwarfs
from DR5 \citep{AdelmanMcCarthy2007} of the SDSS, $-0.22$ dex kpc$^{-1}$
in a sample of horizontal-branch stars \citep{Chen2011} from DR8
\citep{2011ApJS..193...29A}, and $-0.14$ dex kpc$^{-1}$ 
from a sample of kinematically selected thick disk F/G/K dwarfs \citep{Kordopatis2011}. For samples
targeting stars closer to the disk, \citet{Chen2003} measure a slope of $-0.295$ dex kpc$^{-1}$ using open
clusters, \citet{2003BaltA..12..539B} obtain $-0.23$ dex kpc$^{-1}$ for a
sample of thin-disk stars identified by kinematics and asymmetric drift,  \citet{Marsakov2006} find a slope of $-0.29$ dex kpc$^{-1}$
for thin-disk stars selected based on chemistry and orbital parameters, and \citet{Soubiran2007} measure a slope of $-0.31$
dex kpc$^{-1}$ for $|z|<1$ kpc from a sample of red clump giants. 

The structure of the disk is also reflected in detailed abundances, with
stars at greater distances from the plane typically having
higher [$\alpha$/Fe] ratios than stars closer
to the plane (e.g., \citealt{Lee2011,Bovy2012a,Schlesinger2012}). The
nature of the vertical structure continues to be debated, with
some studies suggesting that the thin and thick-disks are distinct
populations, while others report that there is just a single
component with a continuous distribution of properties (e.g., 
\citealt{Ivezic2008, Bovy2012, Bovy2012b, Bovy2012a}). 
Stars at larger distances from the plane have been observed to have
flat or slightly positive radial metallicity gradients 
(\citealt{AllendePrieto2006,Juric2008,2012AJ....144..185C,Cheng2012b}), as have older
stars closer to the plane \citep{Nordstrom2004}. The flattening of the radial gradient with height might be explained by a
transition between the thin- and thick-disk populations; \citet{Juric2008}
estimate that the thick-disk becomes the dominant population at about
1 kpc above the Galactic plane, which is where the radial metallicity
gradients have been observed to be flat.

Models that attempt to explain the properties of the disk can
be better constrained with metallicity information covering a wide
range of the Galactic disk. In particular, the ability of radial
migration models to explain the scatter in the local population depends
on the metallicity distribution function in the inner Galaxy.
Whether radial mixing can heat stars enough to result in a thick-disk
population with the vertical scale height of the Milky Way's thick
disk is the subject of much theoretical dispute (e.g.,
\citealt{Minchev2012c}).
Observational constraints can be provided by comparing the vertical
metallicity and [$\alpha$/M] gradients with populations along the Milky Way
disk to identify the possible origins of such stars if they arrived
by heating.

Measuring chemical abundances in the plane of the disk for a sample covering a large
range of Galactocentric radii is challenging because of the significant
extinction within the disk. The Apache Point Observatory Galactic Evolution
Experiment (APOGEE; \citealt{AllendePrieto2008}) of the Sloan Digital Sky Survey-III (SDSS-III; \citealt{Eisenstein2011}) provides
a unique opportunity to map the chemistry of stars in all Galactic zones,
because it obtains spectra in the near-IR, where the effects of extinction are reduced. APOGEE is a high-resolution ($R\sim 22,500$)
spectrograph that records stellar spectra in the H band between
1.51 and 1.70 $\mu$m using the SDSS 2.5m telescope \citep{Gunn2006}. The APOGEE survey aims to obtain high signal-to-noise ratio
(S/N) spectra
of $10^5$ stars over three years of operation, with the majority of these targets
being red giants. Because of the intrinsic luminosity of giants and
the significant reduction in extinction in the H band relative to
the optical, APOGEE is capable of observing stars directly in the plane
of the Galaxy to large distances. As a result, the survey provides
excellent coverage over a large range of Galactocentric radii and enables
the characterization of stellar abundances across the Galaxy. The
eventual goal is to determine individual elemental abundances for $\sim$
15 different elements for the bulk of the survey stars.

Having a sample of tens of thousands of objects covering the Galaxy
from the bulge to the edge of the disk, at a large range of heights
above the plane, combined with accurate chemical abundances for up to
15 elements, makes APOGEE unique in its ability to study the Milky Way.
Previous studies of the radial gradient used bright tracers that lie in
the plane of the Galaxy (e.g., Cepheids [\citealt{Luck2011a}], HII regions
[\citealt{Balser2011}]), but lacked vertical coverage and had small sample
sizes. Studies with larger sample sizes, such as large-scale surveys like
SEGUE \citep{Cheng2012b} or the GCS \citep{Nordstrom2004}, were done with
observations taken in the optical regime and had limited radial coverage in the plane of 
the Galaxy. The vertical gradient has only been characterized near the solar circle
(e.g., \citealt{Ak2007,Chen2011,Kordopatis2011}), and has not been measured at other locations
in the disk. APOGEE is the first survey that provides
a large sample with excellent spatial coverage of the Galaxy, allowing
the simultaneous determination of radial and vertical gradients across the disk.

In this paper, we present results on mean abundances in the disk using
data from the first year of the APOGEE survey. We restrict our discussion
primarily to the overall metal abundance, although we use the derived
[$\alpha$/M] ratios to define low-[$\alpha$/M] and high-[$\alpha$/M]
subsamples. A complementary paper \citep{Anders:submitted-a} presents results from a
subsample of the first year data for which 3D kinematics can be derived.
Discussion of other elemental abundances is deferred to
future work, as the ability to derive these abundances is still being
developed. 

\section{Data and Sample Selection}

\begin{figure*}[ht!]
\centering
\includegraphics[width=5.2in]{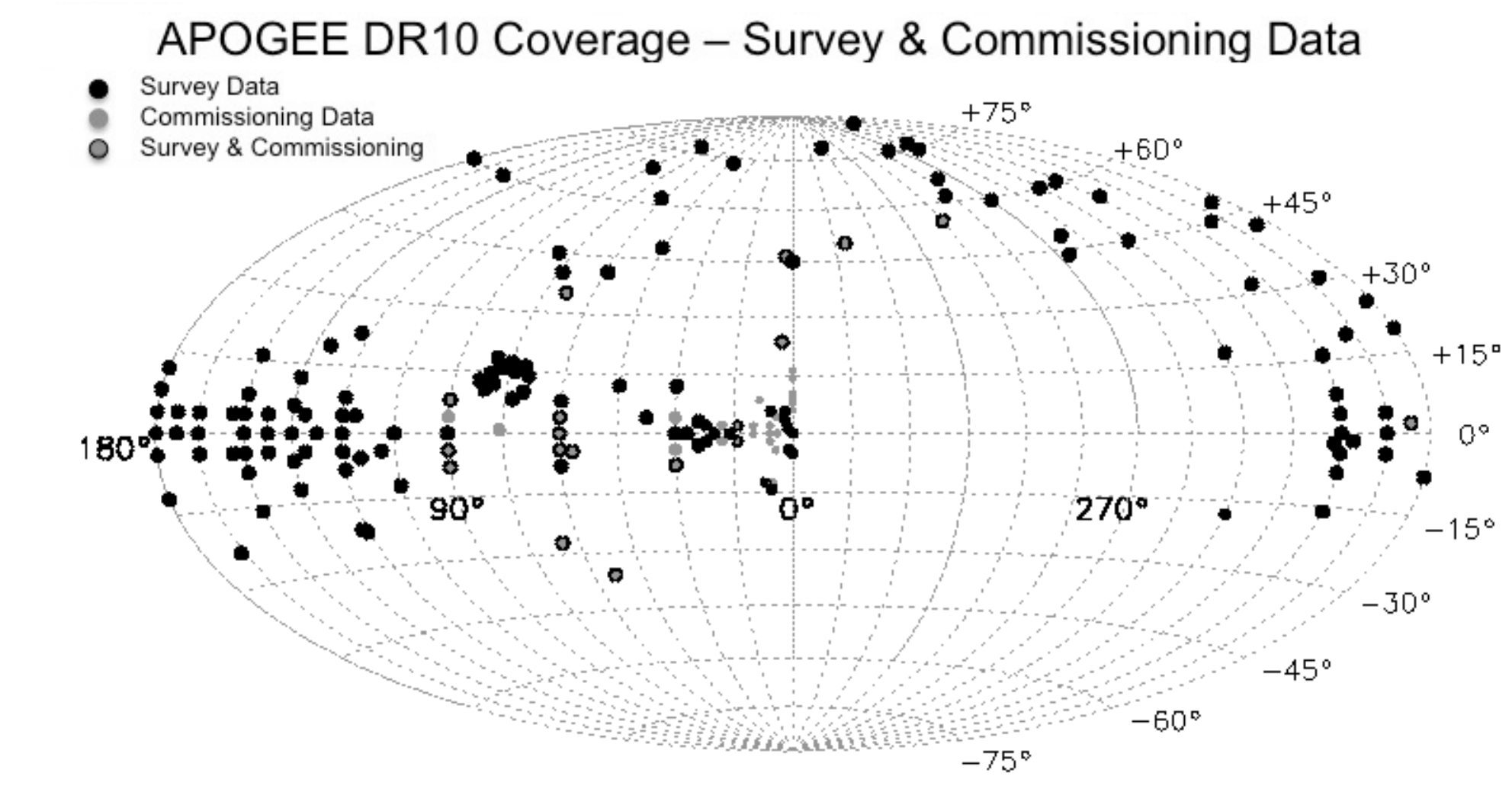}
\caption{The APOGEE footprint for DR10 observations overlaid on Galactic coordinates. The large points are 3$^\circ$ fields, while the smaller points in the bulge are 1$^\circ$ fields.}
\label{survey}
\end{figure*}

Data are taken from the Tenth Data Release of
the Sloan Digital Sky Survey (DR10; \citealt{Ahn2013}), which contains 
stellar spectra and derived stellar parameters for stars observed during 
the first year of SDSS-III/APOGEE. These stars cover a wide range of the 
Galaxy (Figure \ref{survey}), and  span a range of $8<H<13.8$ for primary 
science targets. Target selection is described in
detail in \citet{2013AJ....146...81Z}. To summarize briefly, the sample
is selected from the 2MASS catalog using a dereddened color cut of
${(J-K)}_0 > 0.5$ to remove hotter main sequence stars and to ensure 
that only stars with T$_{\mathrm{eff}}\lesssim 5500~K$
(for which stellar parameters and abundances can be accurately determined)
are selected, but without biasing the sample against metal-poor stars. The
dereddening is accomplished using the RJCE method \citep{Majewski2011},
which uses 2MASS photometry in conjunction with near-IR photometry
from Spitzer/IRAC \citep{Fazio2004} where available or from WISE
\citep{Wright2010}.

APOGEE is a fiber-fed spectrograph that records 300 spectra simultaneously. 
Standard fields include
230 science targets, 35 hot stars to characterize telluric absorption, and 35
sky fibers. Exposures of 500s are obtained in a series of 47 non-destructive
readouts and are taken in pairs, with a 0.5 pixel detector shift between the
two exposures to provide well-sampled spectra across a spectral range that
runs from 1.51 to 1.70 $\mu$m; a typical visit includes four
pairs of exposures. Each field is visited at least three times to identify radial
velocity variables, except for selected bulge fields. 

Data are processed through a standard pipeline that
performs basic calibrations and collapses the data cubes to 2D images, 
extracts spectra, measures radial velocities (RVs), 
and combines spectra from different visits
\citep{Nidever:inprep-b}. Stellar parameters and abundances are derived
by the APOGEE Stellar Parameters and Chemical Abundances Pipeline (ASPCAP; 
\citealt{GarciaPerez:inprep-e}), which determines the best-matching
parameters by searching within a multi-dimensional grid of synthetic
spectra. For DR10, the dimensions that are used to characterize the spectra 
include effective temperature (T$_{\textrm{eff}}$), surface gravity ($\log{\textrm{g}}$), overall metal abundance ([M/H]), 
$\alpha$ element abundances ([$\alpha$/M]), carbon abundance ([C/M]), and nitrogen abundance ([N/M]); the latter three abundances are included as fundamental stellar parameters because of the strong importance
of molecular features from CN, CO, and OH in the region covered by the APOGEE spectrum. The current grid covers T$_{\mathrm{eff}}>3500~$K and $\log \mathrm{g}>0$; stars
with derived parameters that fall close to the grid edges are flagged as
likely to be unreliable.

Additional details on sample selection and
data processing are provided in the APOGEE targeting paper \citep{2013AJ....146...81Z}, the DR10 paper \citep{Ahn2013}, and on the DR10 
web site\footnote{http://www.sdss3.org/dr10}.

The quality of the derived parameters and abundances has been discussed
in \citet{Meszaros2013}, based primarily on observations of stars in
clusters with known abundances. This study found small systematic offsets
in some of the ASPCAP-derived parameters; these corrections have been
applied in the DR10 data release and are used here.  Results for dwarfs
($\log \mathrm{g}>3.8$) were found to be less reliable, and insufficient
calibrators are available for these objects, so we only include stars with $\log
\mathrm{g}<3.8$.

\citet{Meszaros2013} also characterize the uncertainties in derived
parameters and abundances after calibration, and find that [M/H] and [$\alpha$/M] 
typically have a precision of 0.1 dex, uncertainties in temperature are $\sim$
100 K, and uncertainties in $\log \mathrm{g}$ are $\sim$ 0.2 dex. However,
there is some indication that the [$\alpha$/M] results are unreliable for
cooler (T$_{\mathrm{eff}}<4200~$K) stars, as discussed further below. Results for carbon
and nitrogen are considerably less certain and are still a topic of ongoing investigation.

\begin{figure*}[ht!]
\centering
\includegraphics[width=5.2in]{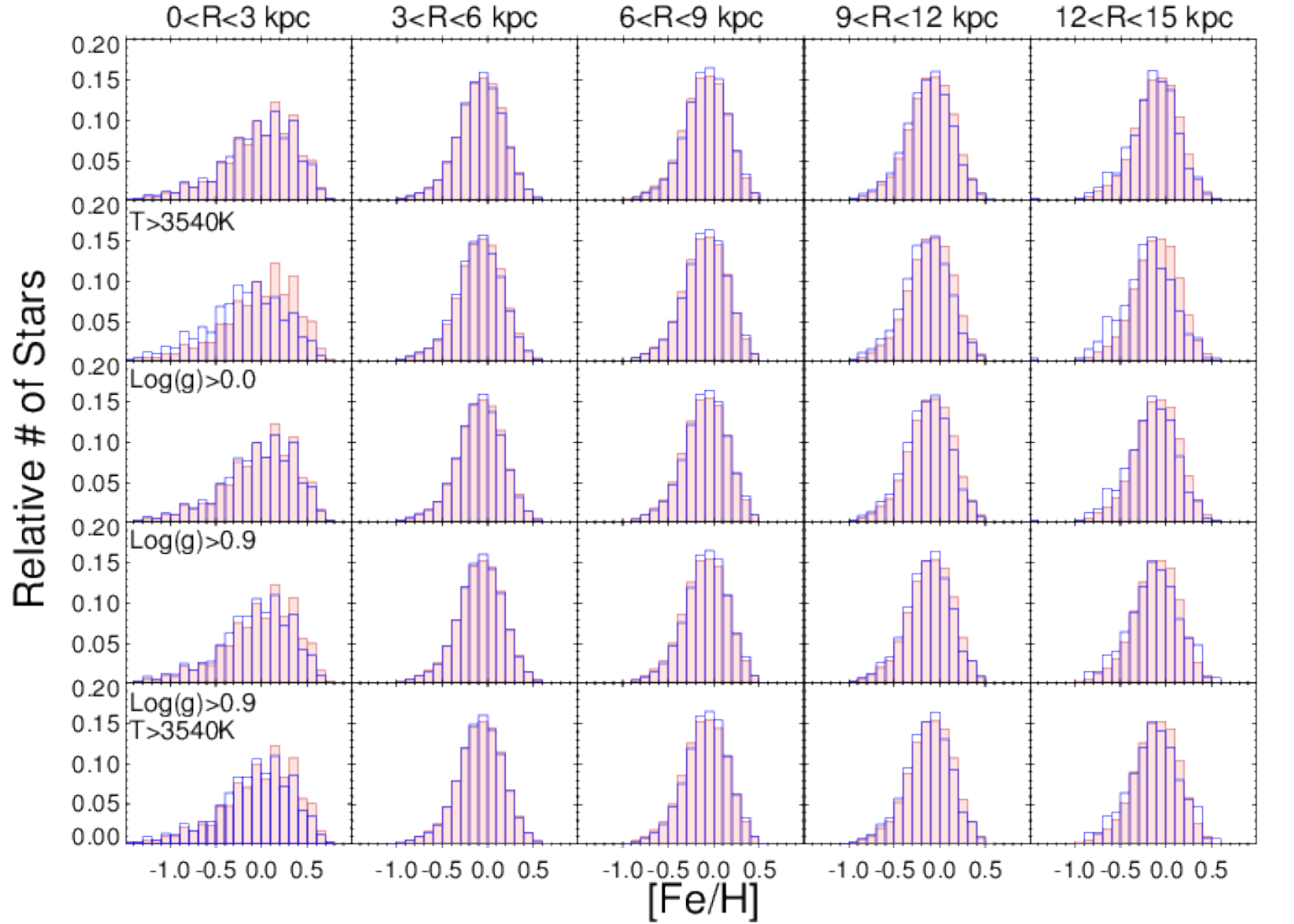}
\caption{The metallicity distribution function (MDF) for the entire input 
TRILEGAL simulation (red filled), versus the MDF of various sample selections 
from the simulations (blue line), in
five different radial bins from left to right. The top row shows target selection only, the second row shows additional
limitation of T$_{\mathrm{eff}}>3540~$K, the third row shows restriction to $\log \mathrm{g}>0$, the fourth row shows $\log \mathrm{g}>0.9$ and T$_{\mathrm{eff}}>3540~$K, and the fifth row shows the combined effects of both the surface gravity and temperature restrictions.  The fourth and fifth rows are the same: the effect of the temperature grid edge on metallicity has been removed with this surface gravity cut.}
\label{trilegalMDF}
\end{figure*}

\subsection{Distances}

Distances for each star are determined from the derived stellar parameters
based on Bayesian statistics, as described in \citet{Hayden:inprep-a},
following methods described by \cite{Burnett2010}; see also \citet{Santiago:submitted-b}. The probability
of all possible distances is calculated for each star given the extinction-corrected
magnitude (using RJCE, as referenced above) and the stellar parameters from ASPCAP,
under the assumption that the parameters are related according to
the theoretical isochrones of the Padova group \citep{Bressan2012}.
The isochrones also account for the relative numbers of stars with
different parameters as determined by the initial mass function and
the timescales for different stages of stellar evolution. Density
distributions of four Galactic components (thin-disk, thick-disk,
triaxial bulge, and halo) are included as priors in computing the
distance probability distribution function, as well as priors on
the age distribution of each component. For each
star, the mean distance, as determined from its distance probability
distribution function, is adopted.

Distances were tested using a set of simulated observations from
TRILEGAL \citep{Girardi2005a}, observations of calibration clusters
\citep{Meszaros2013} of known distance, and observations of Kepler
giants with asteroseismic parameters \citep{Pinsonneault:inprep-c} that
allow independent distance estimates.  Generally, we find
distance errors of $\lesssim 20$\% for individual stars.  The error in $R$
is a function of the error in the distance and the line-of-sight of the observation; errors in 
$R$ are smaller than the error in distance.  Future additions to
the priors, such as the incorporation of the APOGEE selection function and 
3-D extinction maps \citep{Schultheis:inprep-g}, are planned to help improve 
the accuracy of distance estimates.

\subsection{Potential metallicity biases}

Before presenting mean metallicity results, we consider
several potential sources of bias in our observed metallicities:
\begin{itemize}
  \item Bias from sample selection: color cut and magnitude limits.
  \item ASPCAP spectral grid edges: spectral library has no results for T$_{\mathrm{eff}}<3500~$K and $\log \mathrm{g}<0$.
  \item Metallicity bias in derived distances
\end{itemize}

\subsubsection{Biases from sample and ASPCAP selection}

To characterize possible biases, we use a set of simulated observations
from TRILEGAL that were run through our targeting selection scheme
(color cut and magnitude distributions) and the DR10 selection of
fields to determine the effect of target selection on the observed
metallicities. These simulations include different metallicity
distribution functions (MDFs) for different Galactic components, but
not any changes in the MDF of any individual component (i.e., no disk
metallicity gradient). The simulations are still useful for determining whether spatial
variations might be introduced because of sampling effects, since the
typical parameters of stars in different Galactic zones will be different
(e.g., more distant stars will tend to be intrinsically more luminous).  The simulations 
are for fields with $l<90^{\circ}$ (i.e. the inner Galaxy fields), where the effect of the 
ASPCAP grid edge will be the largest on the sample. 

Results are shown in the top row of Figure \ref{trilegalMDF} for
several different Galactocentric radii, in order to illustrate sample selection effects
only; there is little difference in the sampled MDFs
relative to the underlying ones.  

The second row shows the effect of limiting the TRILEGAL sample to T$_{\mathrm{eff}}>3540~$K,
to test the effects of the ASPCAP grid edge on the observed MDF. This constraint clearly leads to a bias against high-metallicity stars in more distant zones, as expected, since the
giant branch is cooler at higher luminosity and metallicity. 
The extent to which such a bias exists in the real data depends on how
many high-metallicity stars actually exist at these distances, but,
since the effect is significant for $\textrm{[M/H]}>0.1$, this effect is likely to be
an issue for the APOGEE survey.  The overall number of stars that
have ASPCAP parameters at the grid edges is small: only 300 stars of
our sample of 20,000. However, most of these objects are in the direction of
the inner Galaxy. Because we cannot derive the stellar parameters for these 
stars, we are not able to determine their distance, but it is not
impossible that most of them are located at small Galactocentric radii.  

The bias against high metallicities stars also appears in the simulations
at large Galactocentric radii, but this is likely due to the fact that
sampled simulations have only been made for $l < 90^{\circ}$, so all 
of the stars in the simulation at large Galactocentric radius are at 
very large distances from the Sun, and, as a result, are of very low
surface gravity and temperature. In the actual APOGEE data, most of
the stars at larger Galactocentric radii are in the direction of the
anti-center, and thus at smaller distances from the Sun, so they are dominated
by stars of higher surface gravity and temperature, for which the bias
is less significant. Furthermore, there are very few stars near the grid
edge in the outer regions.

The third row shows the effect of removing stars with $\log \mathrm{g}<0.0625$, as
also imposed by the ASPCAP grid. This effect is small, because there 
are few of these stars (less than 4\% of the total sample), and surface 
gravity is not as sensitive to metallicity as temperature.

The metallicity bias from the cool temperature limit can be alleviated
by restricting the sample to stars of higher surface gravity, at
the expense of preferentially removing more distant stars from the
sample.  This is demonstrated in the bottom panels of Figure
\ref{trilegalMDF}, which show the MDFs for a restricted sample
with $\log \mathrm{g} >0.9$, both without and with a temperature
cut of T$_{\mathrm{eff}}>3540~$K.  A small bias is present at larger distances, but
it is significantly smaller than for the sample without a surface
gravity cut, and mostly affects stars of higher metallicity
([M/H]$\gtrsim$ 0.3).  This bias is predominantly a concern for the inner
Galaxy, where more metal rich stars may exist in larger numbers.
Existing measurements of bulge stars (e.g., \citealt{Zoccali2008,Gonzalez2013,Ness2013})
suggest that metal-rich stars are actually less prominent
there than they are in the TRILEGAL simulations, so the bias in the actual data may be smaller than is present
in our simulations. Future work \citep{GarciaPerez:inprep-f} will consider the
MDF of the bulge as derived from APOGEE observations.

The simulations suggest that limiting the sample to $\log \mathrm{g} >0.9$
avoids significant bias in the mean metallicities, although this does rest
on the assumption that the MDF of the simulation is representative of the
real Galaxy. In general, however,
a viable test for biases can be made by measuring the extent to
which the MDF (or mean metallicities) change as the sample is restricted
to stars of higher surface gravity. 

\subsubsection{Biases from distance determination}

Our method of distance determination is not expected to have a strong
sensitivity to metallicity, because the dominant parameter that constrains
the distance is the observed surface gravity, and this depends only
weakly on metallicity.  We verified this expectation by recovering distances from
a TRILEGAL simulation where errors in metallicity, surface gravity,
and temperature were added according to our estimates of how accurately
they are measured by APOGEE ($100~$K in T$_{\textrm{eff}}$, 0.1 dex in [M/H], and 0.2 dex
in $\log \mathrm{g}$). No significant trend in distance error is seen with
the input error in metallicity. Instead, as expected, the distance errors are
significantly correlated with the input error in surface gravity.

Another bias could arise from the use of expected density
distributions for each Galactic component in our Galactic model
priors. However, we have explicitly removed any assumption about the MDF
of the different Galactic components from the priors we use for the
distance determination. We have also verified that, qualitatively, the
results are not significantly different when we remove the use of these
priors for the bulk of the disk sample. An exception is for stars in the
direction of the bulge, for which the density distribution of the prior
drives the distances to larger values; while we present results of these
stars in this paper, we restrict most of the discussion to stars with 
$R>3$ kpc.

\section{Results}

To determine mean metallicities, we use
a sample that includes all stars from the APOGEE DR10 release that
were targeted as part of the ``main survey'' (i.e., we did not
include special targets including clusters,
calibration stars, stars observed for ancillary programs, etc.),
had $\log \mathrm{g}>0$, have S/N$>80$ per pixel in the combined
spectrum, and are not flagged as bad by the ASPCAP pipeline (which
includes flagging stars near the grid edges as bad; see the DR10
web documentation \footnote{http://www.sdss3.org/dr10/irspec/} for more
details). In addition, we consider only stars less than 3 kpc from the
Galactic mid-plane ($|z|$). These selection criteria yield a sample of 19,662 giants.

\begin{figure}[ht!]
\centering
\includegraphics[width=3.2in]{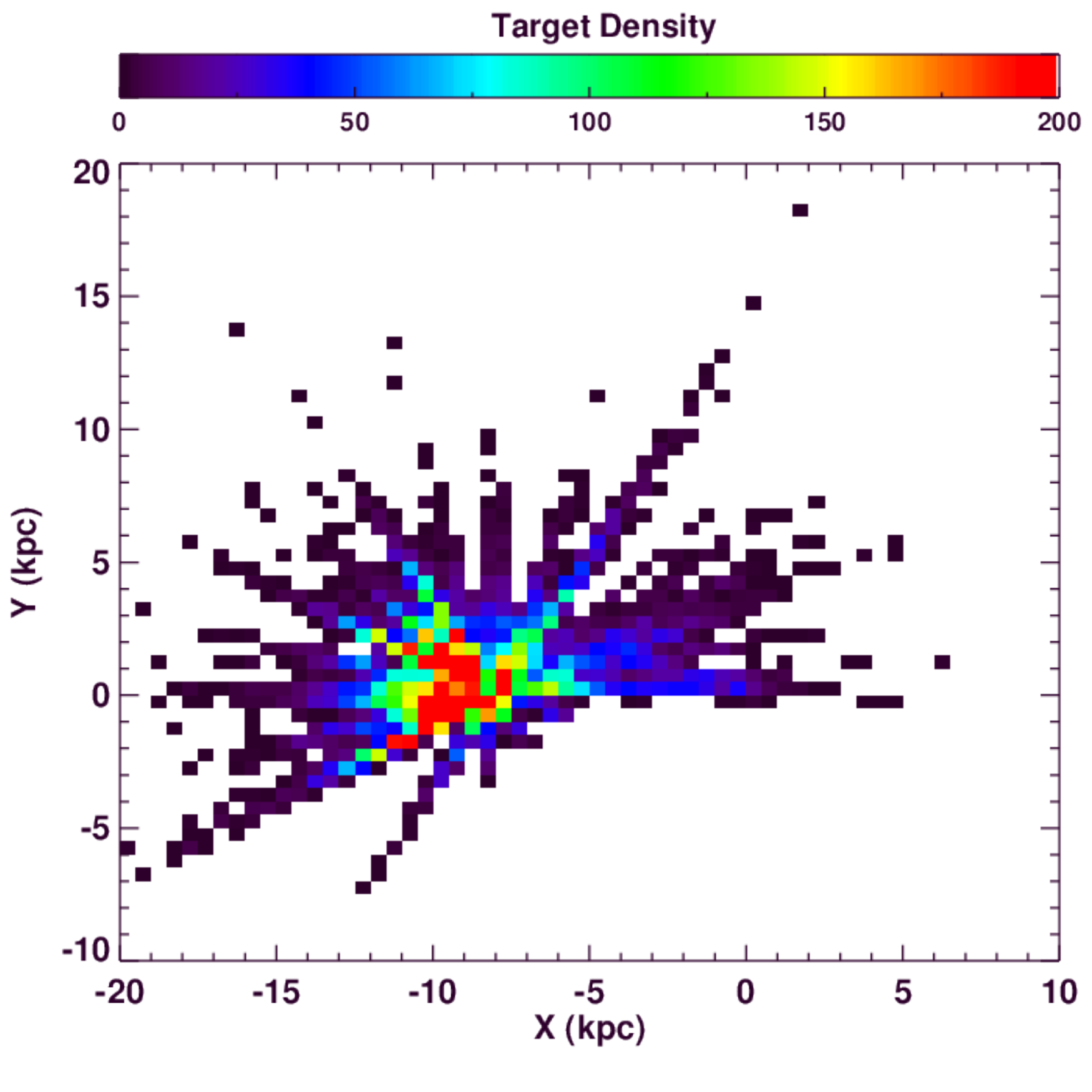}
\includegraphics[width=3.2in]{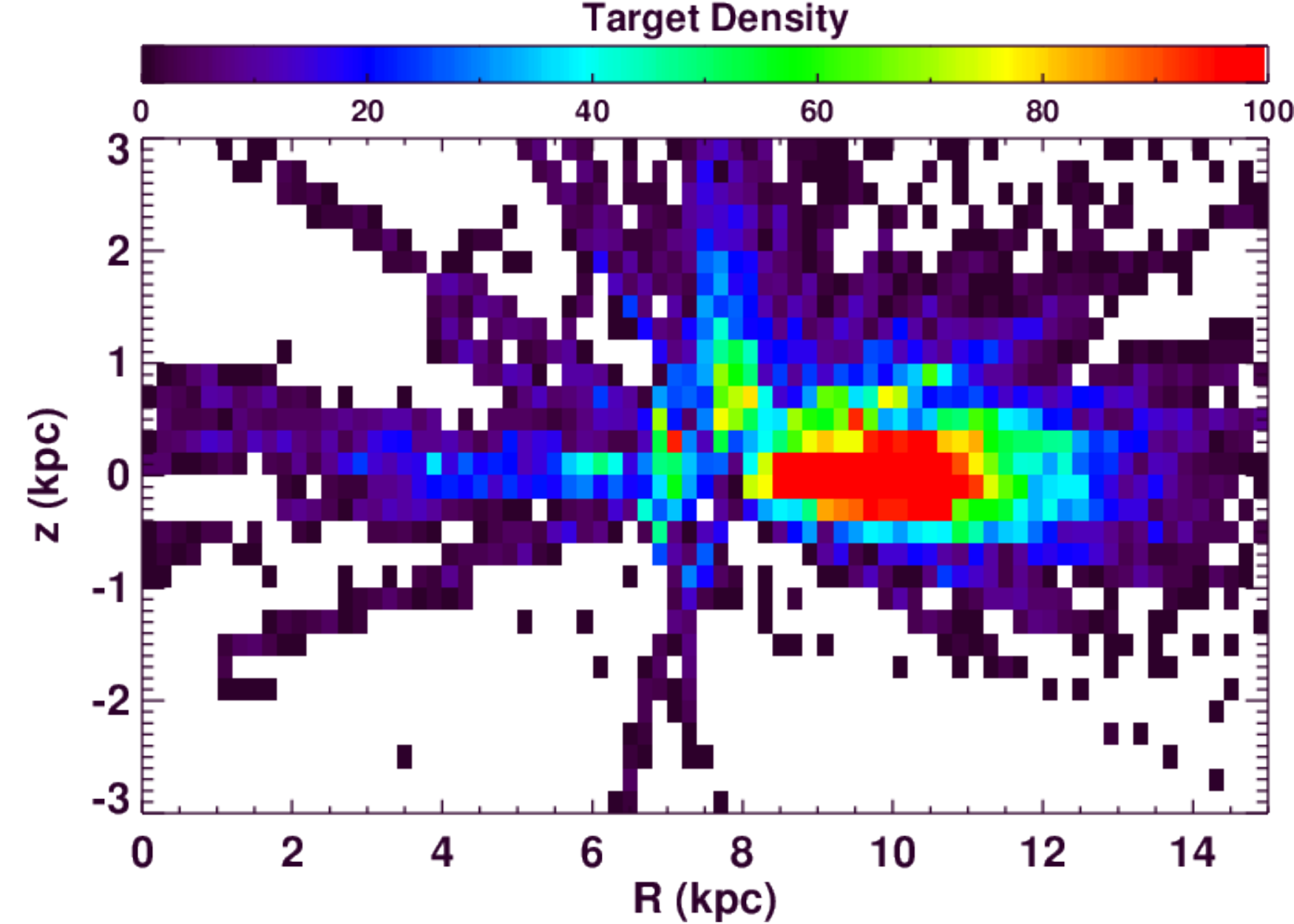} 
\caption{\textit{Top}: A face-on view showing the stellar density in the APOGEE
DR10 sample for stars with $|z|<2$ kpc and $\log \mathrm{g}>0.9$.  The 
Galactic Center is at (0,0), while the Sun is located at ($-8$,0 kpc). 
\textit{Bottom:} The spatial density of targets in the $R-z$ plane.  There are more targets above than below the plane, and more targets in the anti-center direction.}
\label{map}
\end{figure}

Figure \ref{map} displays the observed number of stars in different Galactic 
zones for this sample. The top panel shows a face-on view of the 
Galactic disk, while the bottom panel shows an edge-on view in Galactocentric 
radius and distance from plane. The year 1 sample contains stars covering
$0 < R < 15$ kpc and $-3 < z < 3$ kpc (mostly at $z>-1$) from the plane.  There
are significantly more stars in the anti-center direction because more fields
were completed in that direction during the first year of observation.

\begin{figure}[ht!]
\centering
\includegraphics[width=3.2in]{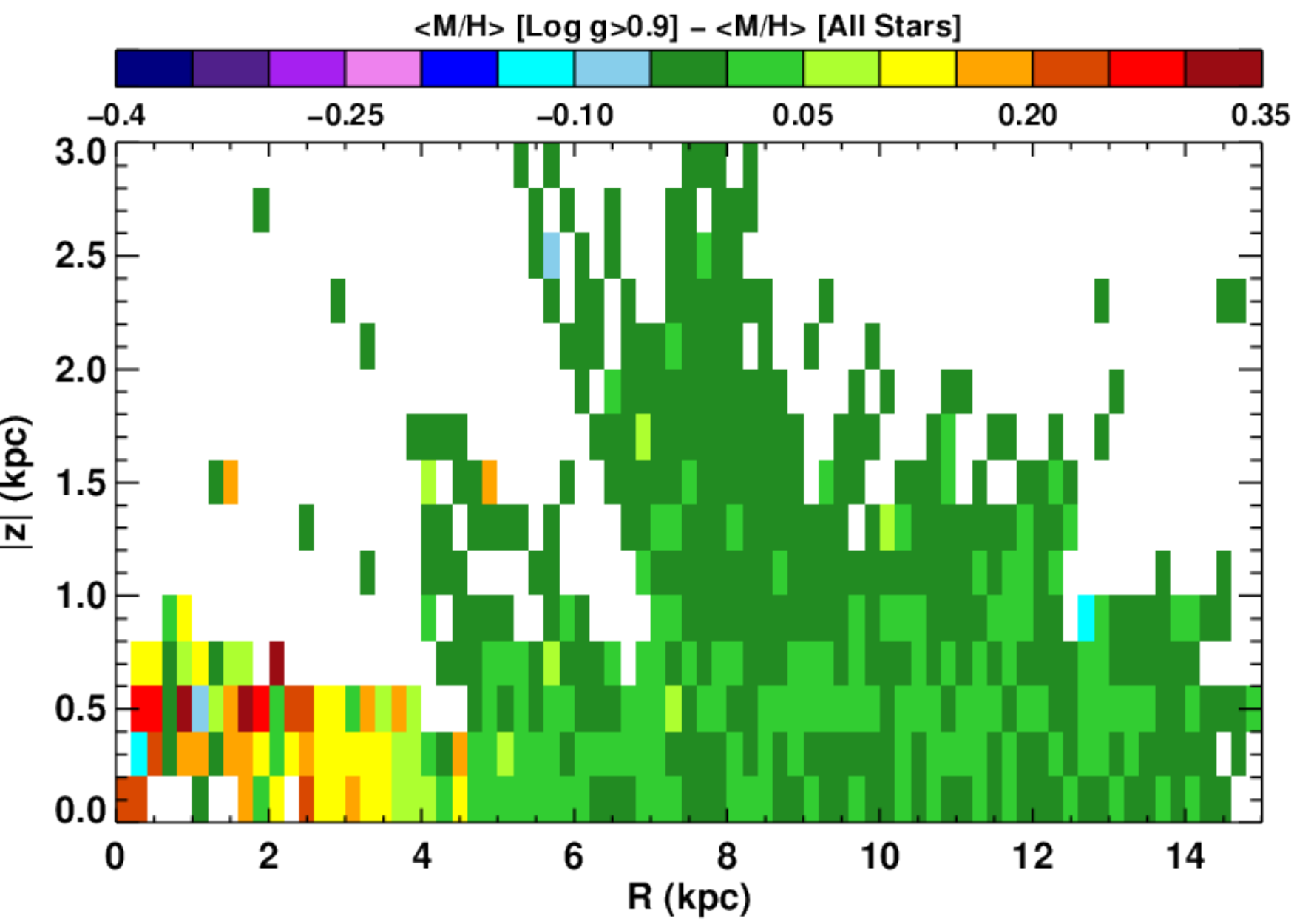}
\includegraphics[width=3.2in]{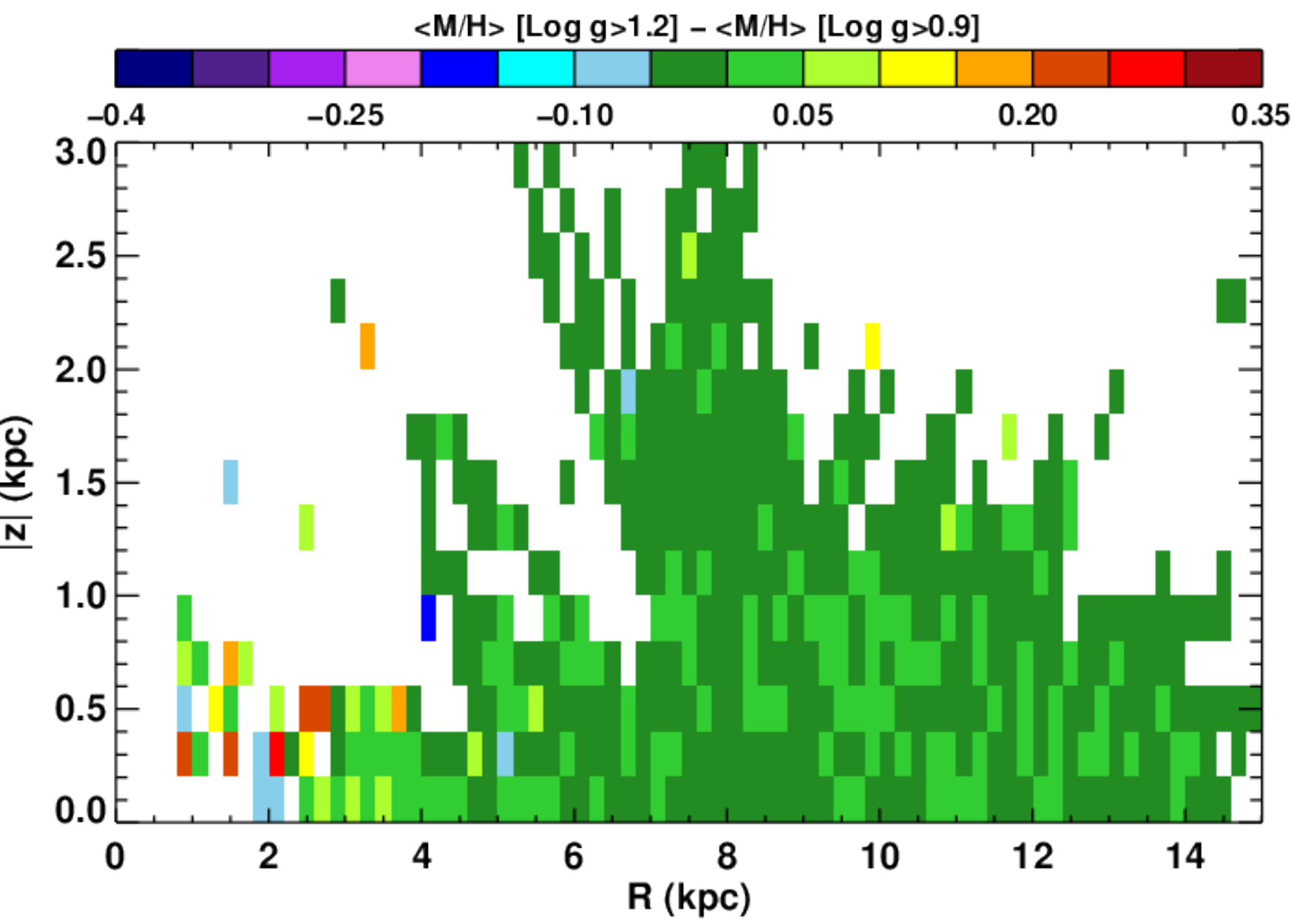}
\caption{\textit{Top:} Difference in mean metallicity of the full sample compared to that of the $\log{\textrm{g}} >0.9$ sample. The difference in mean metallicity is large in the inner Galaxy ($R<4$ kpc), which is where the bias from the ASPCAP grid edge impacts our results.  \textit{Bottom:} same, but between $\log{\textrm{g}}> 0.9$ and $\log{\textrm{g}}>1.2$ samples. The difference in metallicity between this samples is small, and generally less than 0.1 dex, implying that a more stringent surface gravity cut is not required to account for the metallicity bias introduced by the restriction to T$_{\textrm{eff}}>3540~$K.}
\label{fehdiff09}
\end{figure}

To consider whether we are likely to have metallicity biases in the
sample, the top panel of Figure \ref{fehdiff09} presents the difference
between the mean metallicity of stars in the full sample to the mean
metallicity of stars in a sample with $\log{\textrm{g}} >0.9$, and the bottom panel
shows the ratio of a $\log{\textrm{g}}>0.9$ to $\log{\textrm{g}}>1.2$ sample, as a function
of Galactocentric radius and distance from the plane. These results suggest
that metallicity biases are significant only in the inner Galaxy,
at $R<5$ kpc for the full sample, and that a sample with $\log{\textrm{g}}>0.9$ is not significantly
affected by biases for $R\gtrsim 3$ kpc.  We adopt the $\log{\textrm{g}}>0.9$ cut for this paper, and 
future references to the dataset include this cut unless otherwise noted.

\begin{figure*}[ht!]
\centering
\includegraphics[width=6.6in]{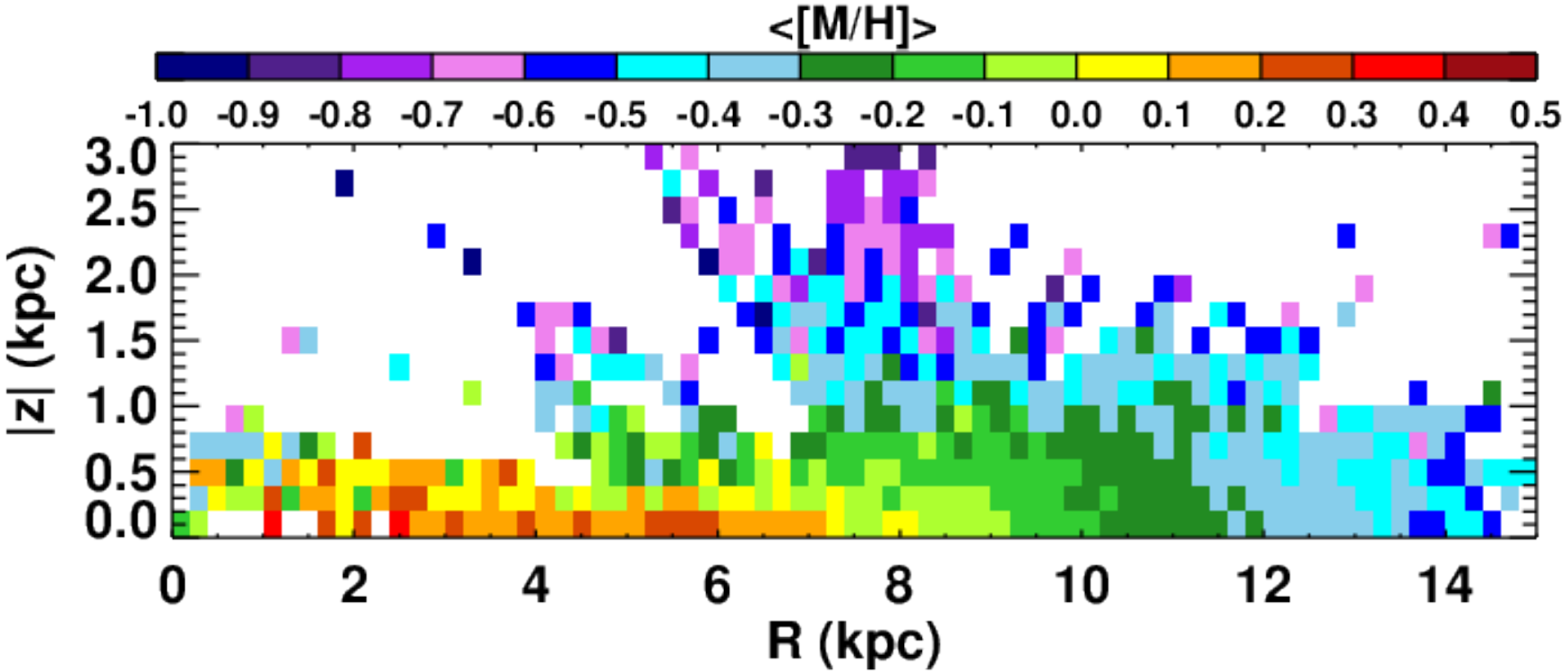}
\includegraphics[width=6.6in]{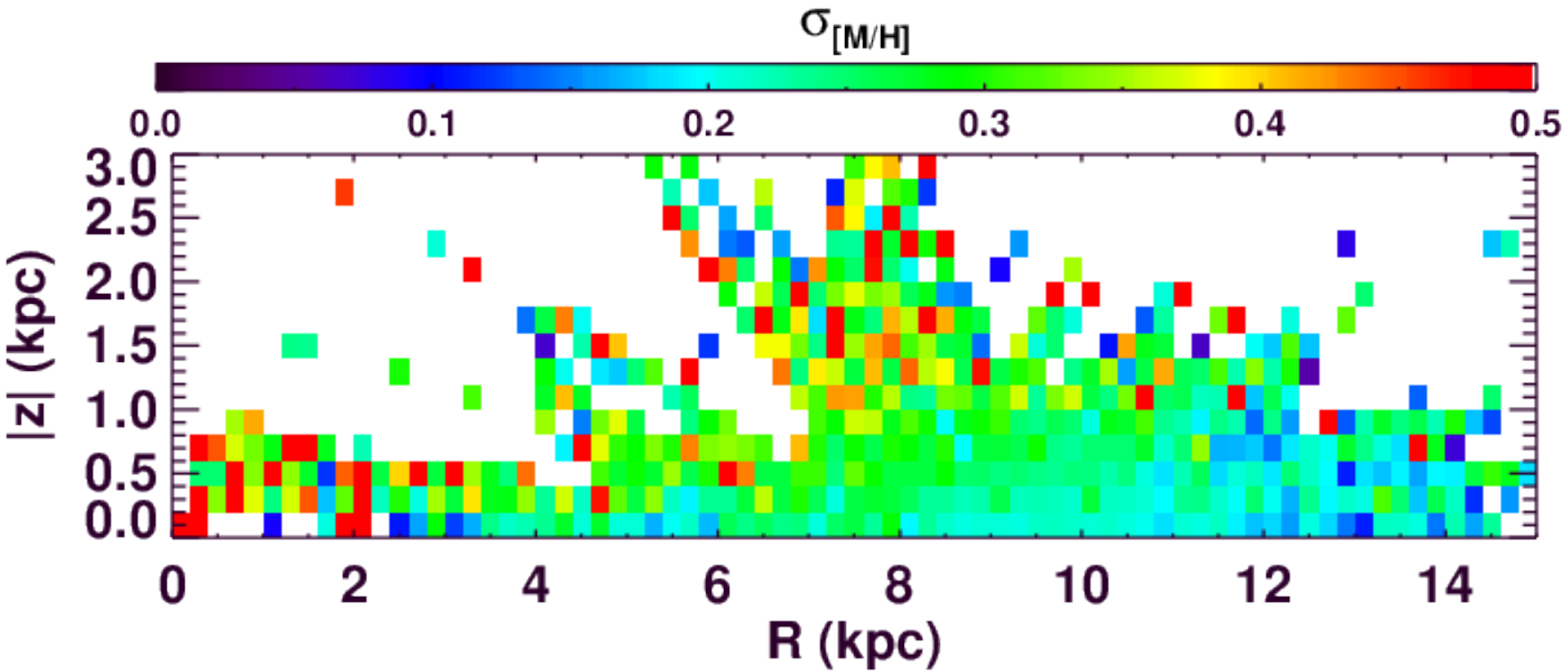} 
\includegraphics[width=6.6in]{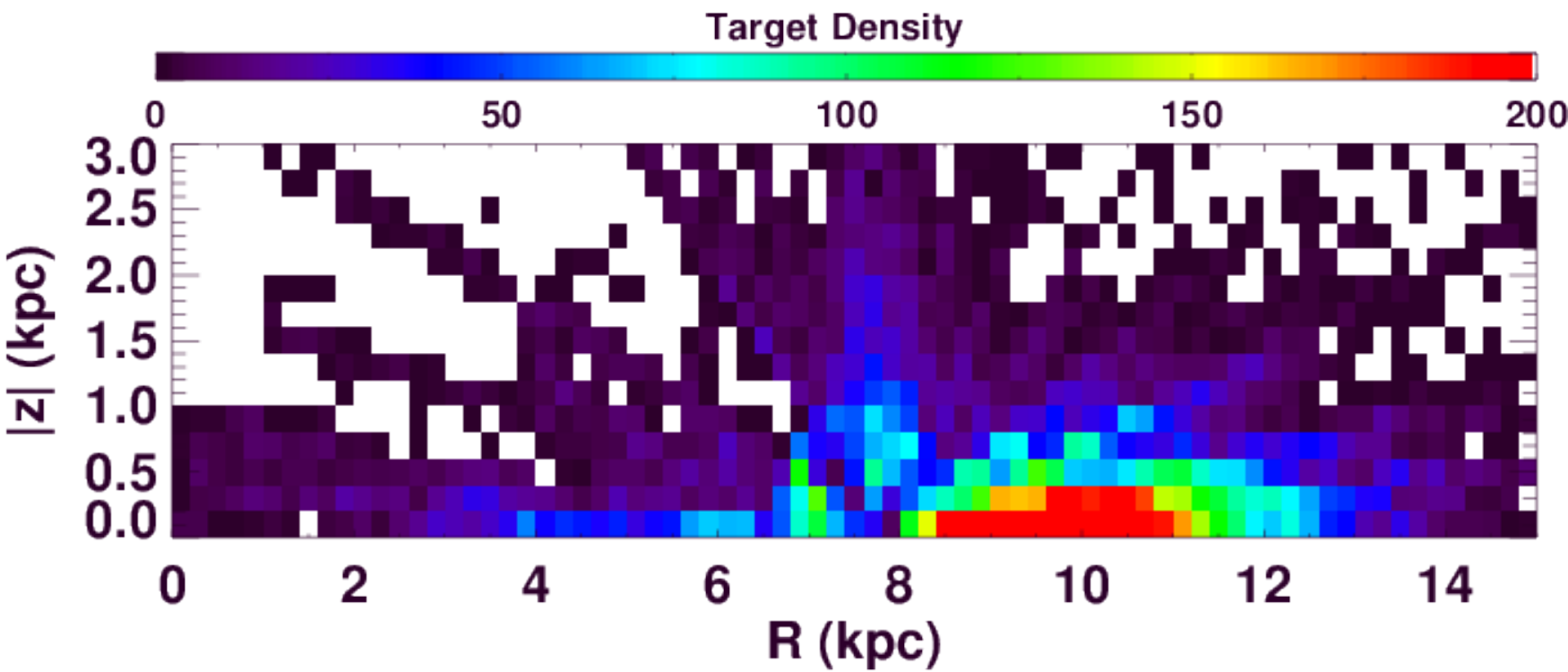} 
\caption{\textit{Top}: The mean metallicity as a function of $R$ and $|z|$ for our sample of stars with $\log \mathrm{g}>0.9$, in zones of $0.2\times0.2$ kpc; a minimum of 4 stars is required to show a bin. The mean metallicity is roughly constant in the inner Galaxy close to the plane, and a negative gradient is clearly seen for populations with $R\gtrsim 6$ kpc. The mean metallicity also decreases rapidly with height about the plane, and the radial gradients are less pronounced at larger heights about the plane. \textit{Middle}: The standard deviation in metallicity at different locations in the Galaxy. The standard deviation at each location is much larger than the errors in metallicity for a single star. \textit{Bottom:} The number of stars in each zone.}
\label{samples}
\end{figure*}

\subsection{Mean metallicity maps}

Our main result is shown in Figure \ref{samples}.
The top panel shows the mean metallicity
of the $\log{\textrm{g}} >0.9$ sample in 0.2x0.2 kpc bins
in Galactocentric radius and distance from the Galactic plane.  In
all zones, there is a substantial metallicity spread well in
excess of the expected abundance accuracy of 0.1 dex for an individual star; this is
quantified in the central panel, which presents the standard deviation 
in each bin. The bottom panel gives the total number of stars in each bin.

Several features are immediately evident from Figure \ref{samples}:
\begin{itemize}
\item The radial mean metallicity variation decreases as one moves away from the plane (see Section \ref{sect:radial}).
\item In the plane, while the overall radial gradient is negative, it appears
to flatten in the inner regions (see Section \ref{sect:radial}).
\item There are significant vertical metallicity gradients (see Section \ref{sect:vertical}).
\item The vertical gradient is steeper in the inner regions of the Galaxy
than in the outer regions. (see Section \ref{sect:vertical}).
\item The spread in metallicity is everywhere larger than the uncertainty
in the abundance determination; in the plane, the variation appears to be larger at 
smaller Galactocentric radius, and it also appears to increase as one moves
above the plane.
\end{itemize}

APOGEE's red giant sample likely includes stars of a wide range of
ages, so the observed metallicity map is almost certainly averaging over
multiple populations of stars. In the case of a constant star
formation rate where the distribution in stellar ages in the Galaxy
is flat, isochrones suggest that the median age of giants
is several Gyr, but depends on the luminosity of the giants:
for red clump stars, the median age is $\sim$ 2 Gyr \citep{Girardi2001},
while for more luminous giants it is somewhat older. This
implies that the mean metallicity is weighted towards larger ages
for more distant zones -- even if the star formation history was
the same in different regions -- and this could influence the interpretation
of metallicity gradients from the entire sample. Variations in 
star formation history would further complicate the interpretation.

\begin{figure}[ht!]
\centering
\includegraphics[width=3.4in]{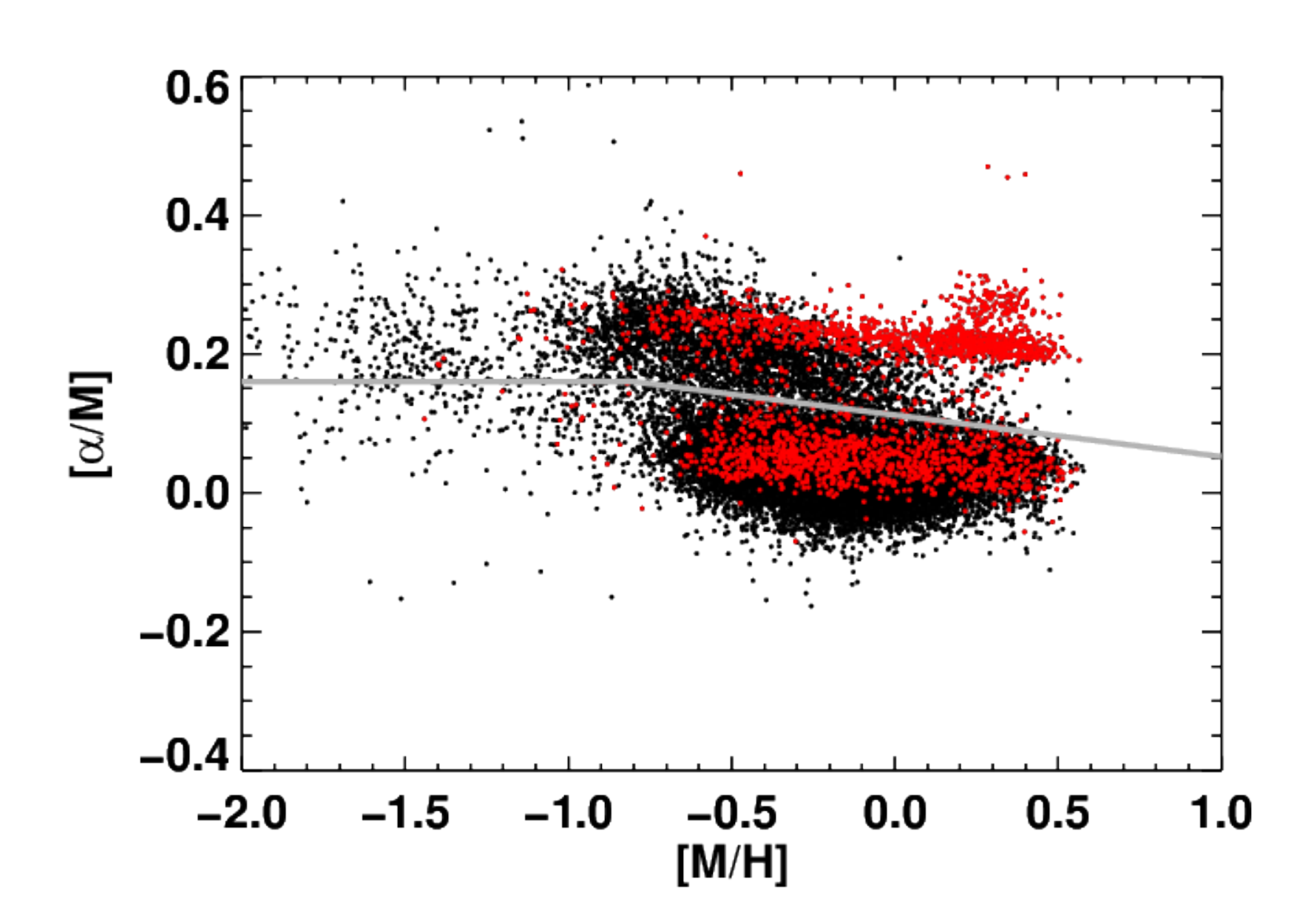}
\includegraphics[width=3.4in]{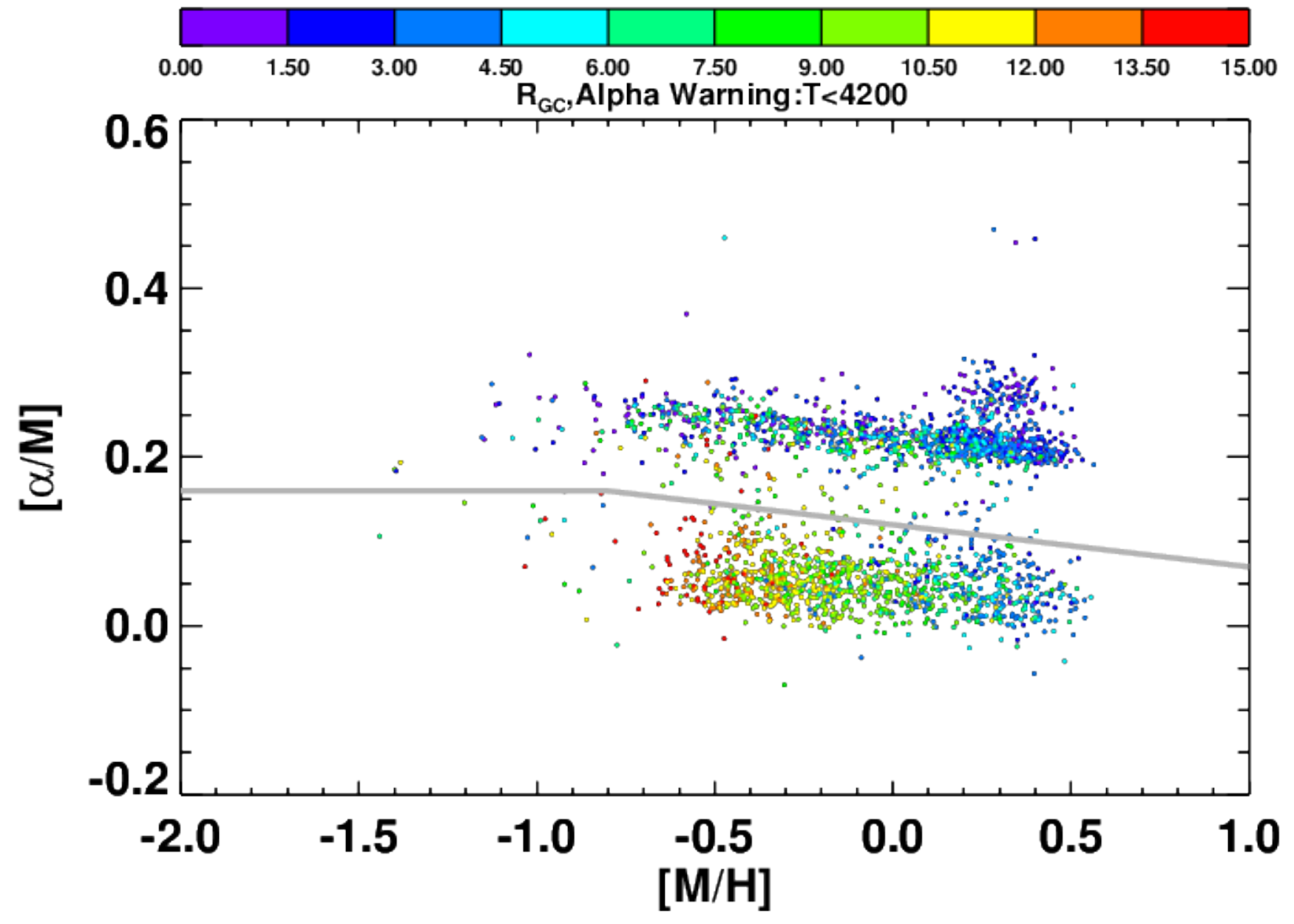}
\includegraphics[width=3.4in]{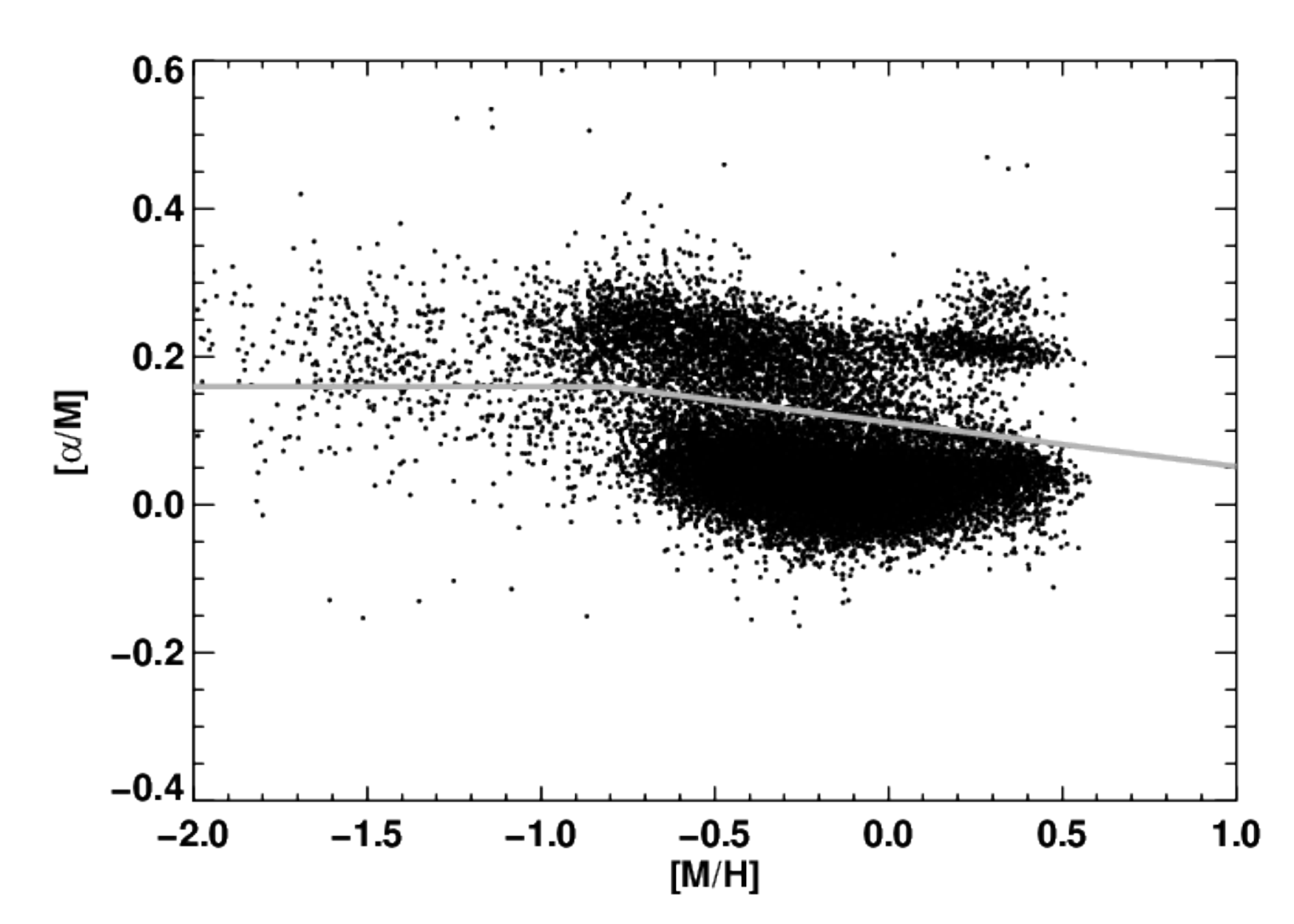}
\caption{\textit{Top}: The APOGEE DR10 sample in [$\alpha$/M] vs
[M/H] space.  Stars with potentially inaccurate [$\alpha$/M]
abundance determinations (T$_{\textrm{eff}}<4200~$K) are shown in red. 
\textit{Middle}: the [M/H] vs [$\alpha$/M] plot for the stars
with T$_{\mathrm{eff}}<4200~$K, color coded by their Galactocentric radius.
The bimodality is still present even for stars in the
same areas of the Galaxy, suggesting that it is likely spurious.
\textit{Bottom}: the [M/H] vs [$\alpha$/M] relation for stars
with T$_{\mathrm{eff}}>4200~$K. In all panels, the grey line shows our 
separation of low- and high-[$\alpha$/M] abundance populations. }
\label{alphacorrection}
\end{figure}

\subsection{[$\alpha$/M] Subsamples}

As a first cut at separating the sample by age, we adopt [$\alpha$/M]
as a potential age discriminator \citep{Lee2011, Bensby2011, Cheng2012a,Bovy2012,
Bovy2012b, Bovy2012a}, where stars with higher [$\alpha$/M] are
more likely to be older stars. Unfortunately, the state of
the ASPCAP pipeline used for the DR10 results has some potentially significant uncertainties with the 
determination of [$\alpha$/M]. This is demonstrated in the top panel of Figure
\ref{alphacorrection}, which shows [$\alpha$/M] as a function of
[M/H] for the APOGEE sample.  Stars with an ASPCAP warning on [$\alpha$/M]
are shown in red, and include all stars with T$_{\mathrm{eff}}<4200~$K. The reason
for the warning flag is apparent from the Figure: the cooler stars
show a strong, narrow bimodal distribution in [$\alpha$/M] that is
not seen at higher temperatures and is not seen by other more local
studies. While one might imagine that this
could potentially be an astrophysical effect, since the cooler stars
are more luminous and thus more likely to be located at greater distances,
we find that the bimodality persists even within stars in the same
Galactic zone (middle panel of Figure \ref{alphacorrection}). When the stars with T$_{\textrm{eff}}<4200~$K are removed (bottom panel of Figure \ref{alphacorrection}), 
the observed [$\alpha$/M] distribution remains bimodal but 
with significant scatter, and resembles results obtained with smaller
samples from the solar neighborhood (e.g., \citealt{Adibekyan2012}). We defer
interpretation of this (e.g., whether it is likely to be a selection
effect or intrinsic to the Galaxy) to a later paper. 

For the present purposes, we split the sample into low- and
high-[$\alpha$/M] samples, along the line shown in  Figure
\ref{alphacorrection}.  This split is similar to that used by
\citet{Lee2011} for their SEGUE sample.  We work with two high-[$\alpha$/M] samples, one including the stars with T$_{\mathrm{eff}}<4200~$K and
one excluding them; while the [$\alpha$/M] results for the cooler stars
are certainly suspect, it is not impossible that they can still distinguish
between lower- and higher-[$\alpha$/M] stars. These cooler giants are
critical for probing to larger distances, since they are intrinsically
more luminous; certainly, however, results from this sample must be
treated cautiously.

The subsamples contain 15,164 giants with [$\alpha$/M] abundances near
solar, and 4,498 [$\alpha$/M]-enhanced giants, if we include stars of
all temperatures. If we remove stars with T$_{\mathrm{eff}}<4200$, we are left
with sample of 14,150 low-[$\alpha$/M] and 3,374 high-[$\alpha$/M] stars;
the temperature cut affects the high-[$\alpha$/M] sample significantly more than
the low one. In subsequent plots, we track three different [$\alpha$/M]
subsamples: low-[$\alpha$/M], all high-[$\alpha$/M], and high-[$\alpha$/M]
with T$_{\mathrm{eff}}>4200$.

\begin{figure}[ht!]
\centering
\includegraphics[width=3.2in]{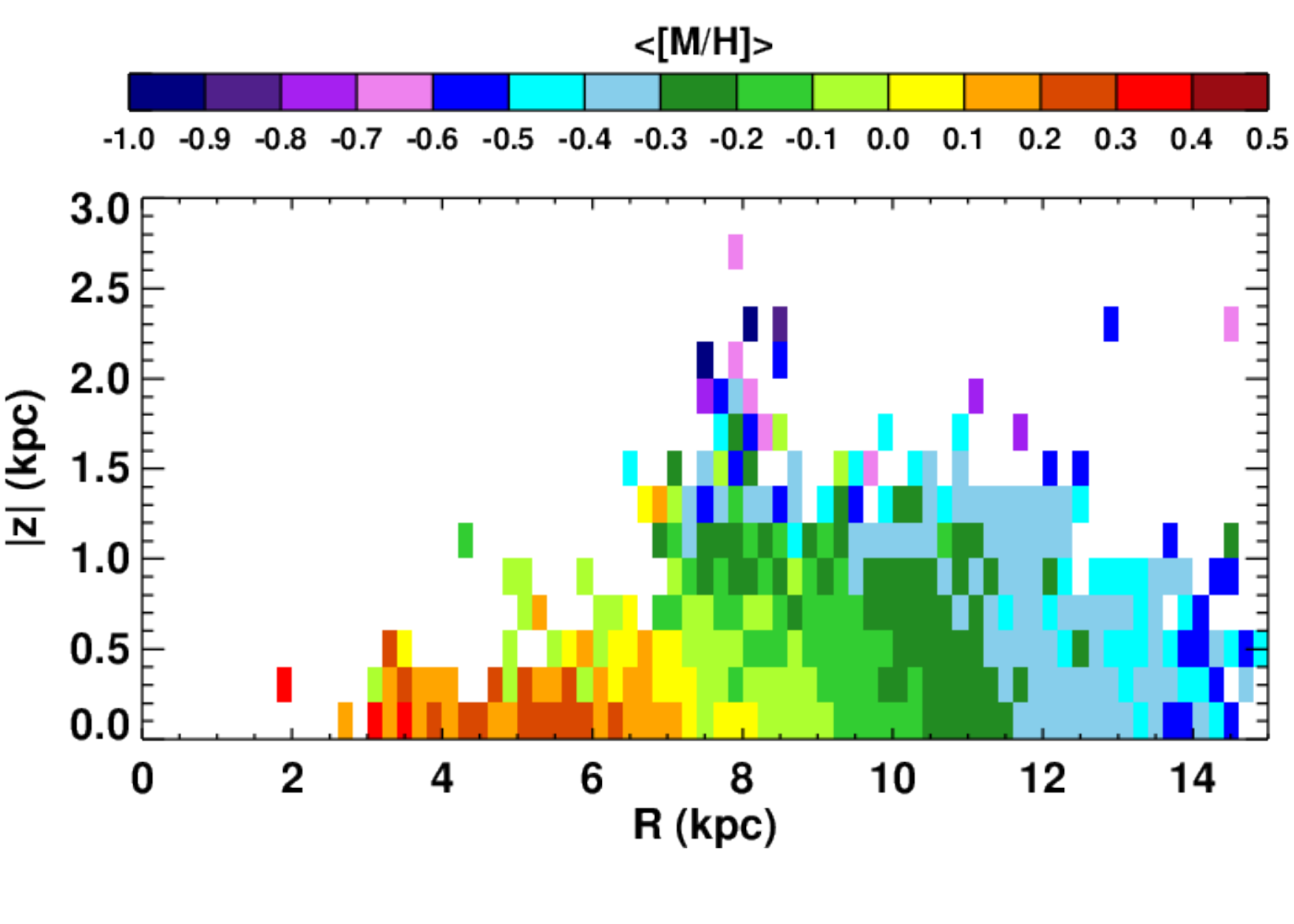}
\includegraphics[width=3.2in]{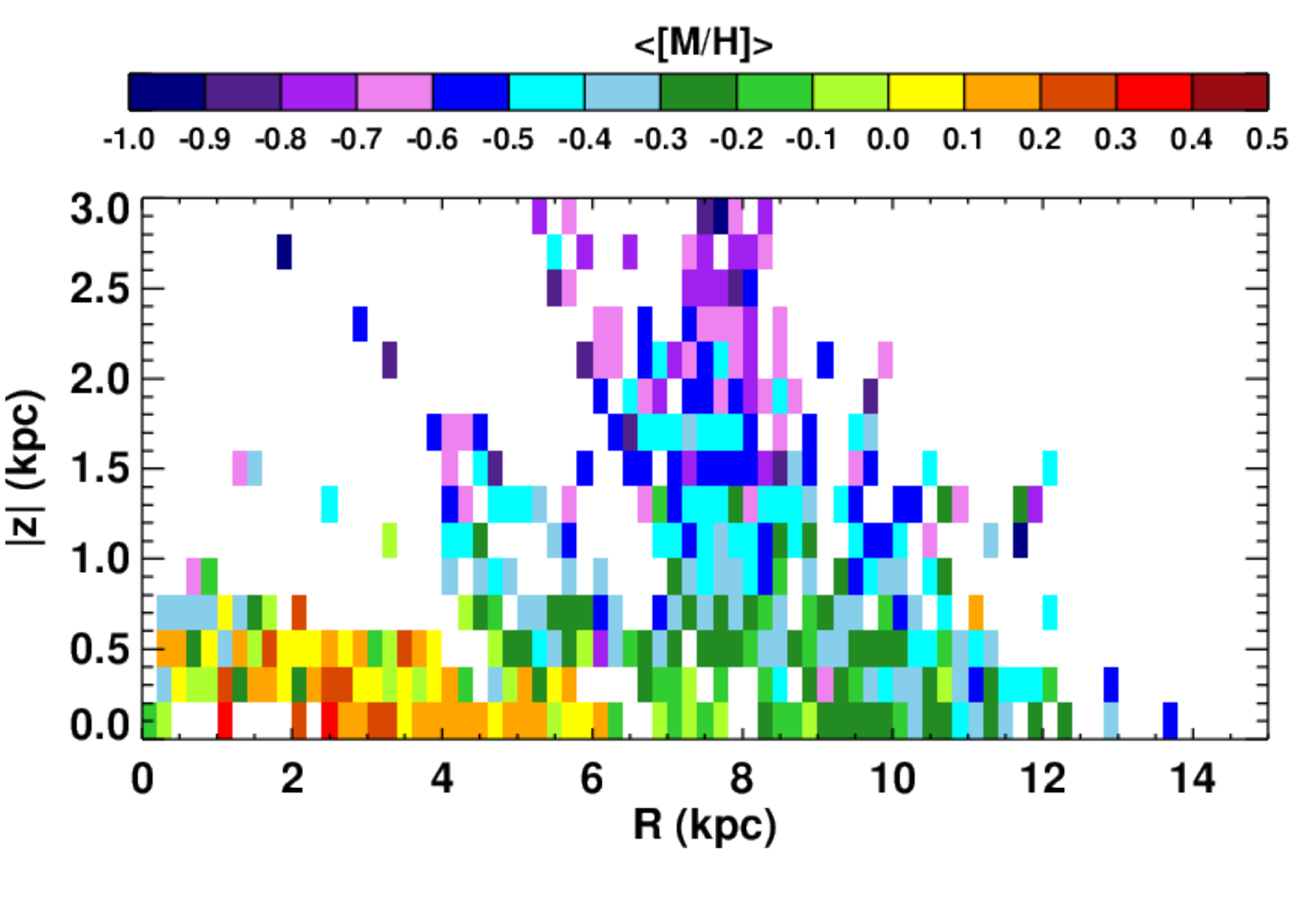}
\includegraphics[width=3.2in]{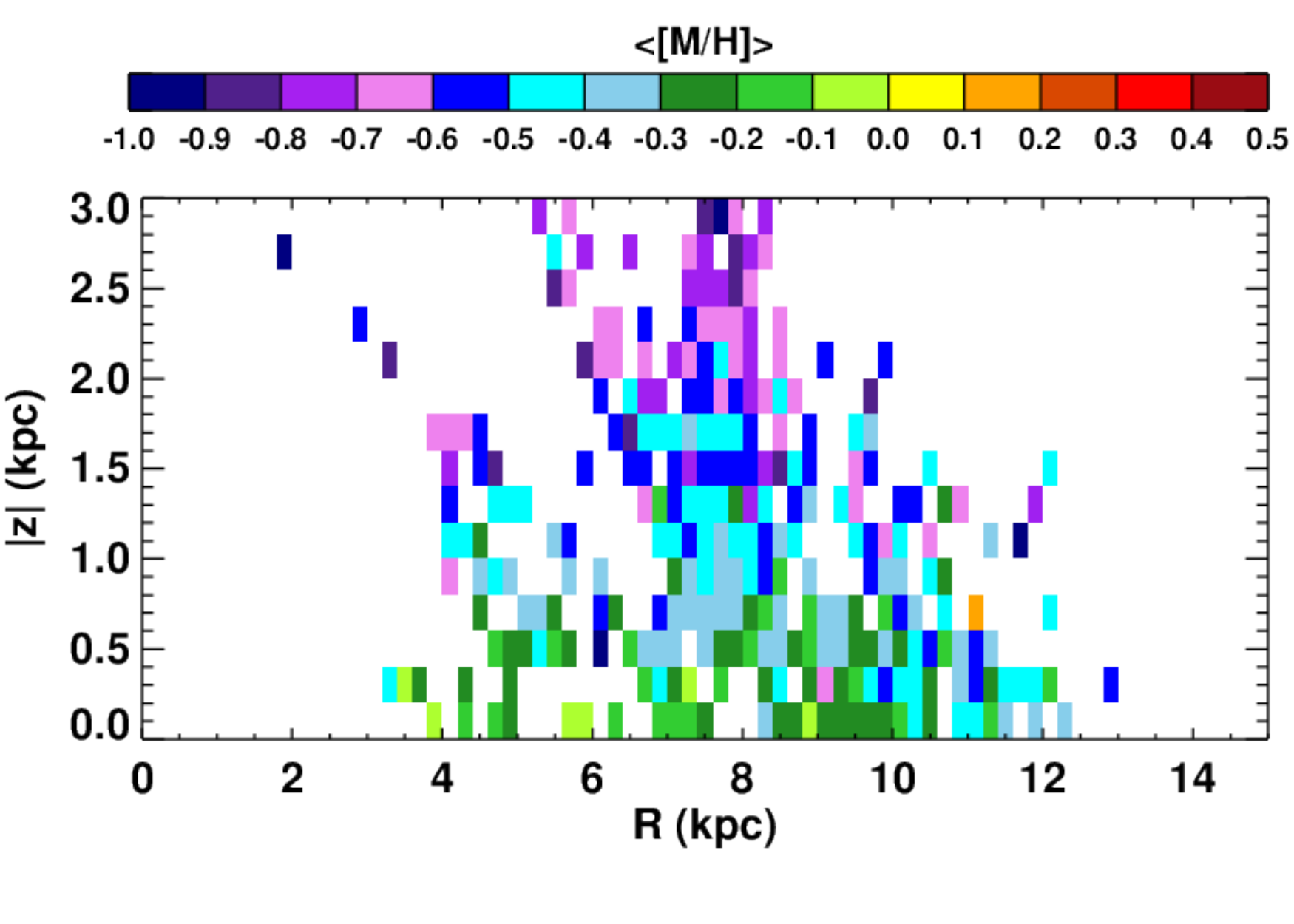}
\caption{\textit{Top}: The mean metallicity as a function of R and
$|z|$ for low-[$\alpha$/M] populations and with $\log \mathrm{g}>0.9$.
The mean metallicity is higher in the plane of the Galaxy than it
was for the entire sample, and the radial gradient is flatter in
the inner Galaxy.
\textit{Middle}: The mean metallicity as a function of R
and \az\s for high-[$\alpha$/M] populations and with $\log
\mathrm{g}>0.9$.  The mean metallicity is 0.2-0.3 dex lower in the
inner Galaxy than for the stars with solar [$\alpha$/M] abundance.
\textit{Bottom}: The mean metallicity as a function of R and \az\s
for high-[$\alpha$/M] populations with $\log \mathrm{g}>0.9$, but
stars that have flagged [$\alpha$/M] abundance measurements 
(stars with T$_{\textrm{eff}}<4200~$K) are removed.  
We lose most stars in
the inner Galaxy, but the mean metallicity in the rest of the Galaxy
is relatively unchanged.
}
\label{lowandhigh}
\end{figure}

Figure \ref{lowandhigh} presents the mean metallicity maps for the different
[$\alpha$/M] subsamples: the upper panel contains the low-[$\alpha$/M]
stars, while the middle and bottom panels show the high-[$\alpha$/M]
stars without and with the low temperature limit, respectively.  The high-[$\alpha$/M] stars are clearly more metal poor, and extend significantly
higher above the mid-plane of the Galaxy; there are very few low-[$\alpha$/M]
stars at $|z|>2$ kpc.  Lower-[$\alpha$/M] stars
extend farther out in Galactocentric radius; there are very few high-[$\alpha$/M] stars beyond $R\gtrsim 12$ kpc. There is a suggestion that
the metallicity gradient for the low-[$\alpha$/M] stars flattens in
the inner regions of the disk. In the outer Galaxy, the gradient in
high-[$\alpha$/M] stars is significantly flatter than that in low-[$\alpha$/M] 
stars.

\subsection{Quantitative gradients}

In order to quantify the observed gradients we measure the vertical and
radial components separately in several different radial and vertical bins.
We first determine the vertical gradient,
and then measure the radial gradient in a series of \az\s bins after
correcting the stars to a common height $\bar{|z|}$, because the vertical
gradient is significantly steeper than the radial one.

\begin{figure}[h!]
\centering
\includegraphics[width=3.2in]{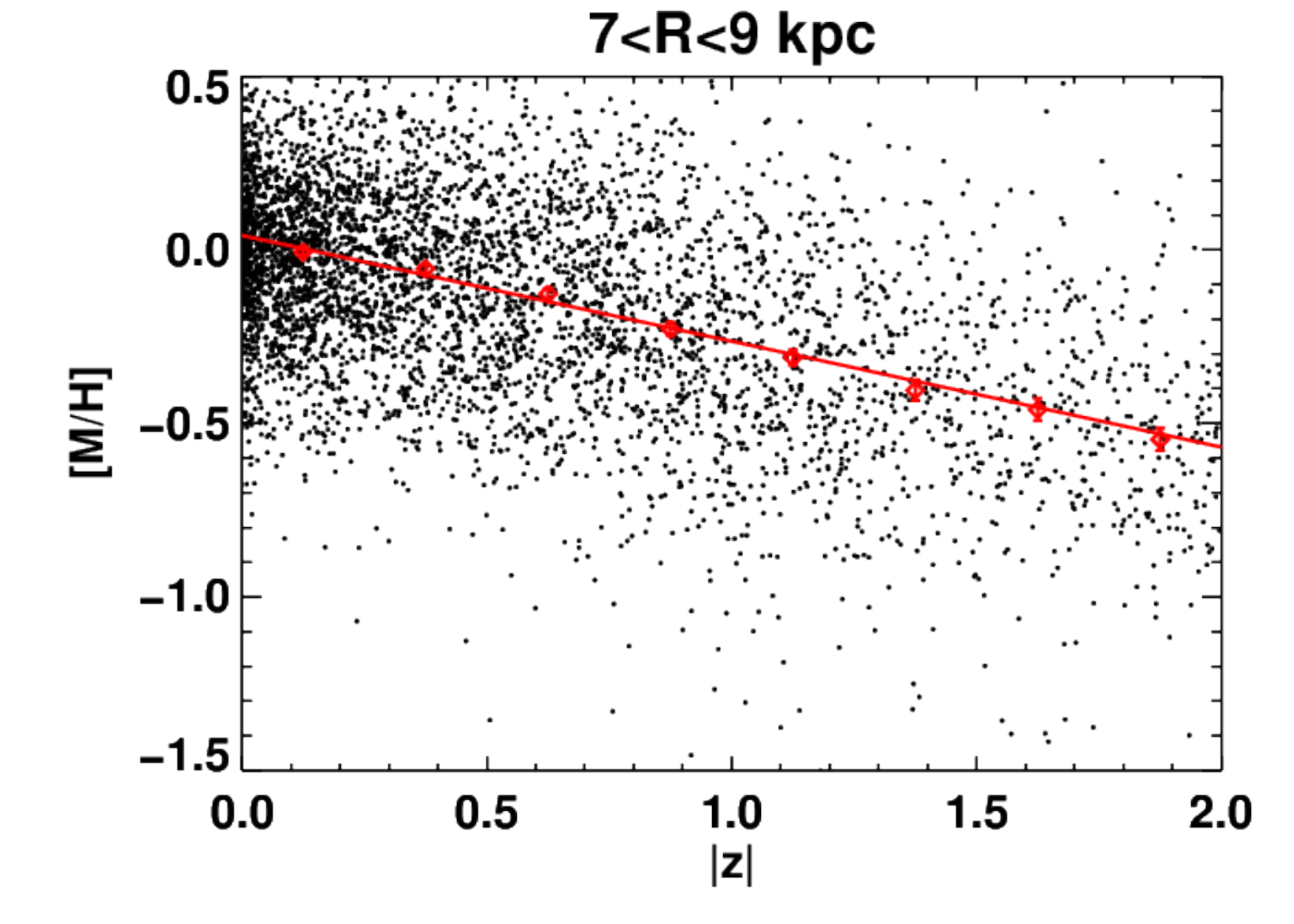}
\caption{The vertical abundance gradient for the entire sample for stars at the solar circle, with the red diamonds representing the median metallicity in each bin.  The slope and intercept of the fit are reported in Table \ref{vertgrad}.  The vertical gradient is smoothly varying over the range of our sample, and no discontinuity (e.g., between a thick and thin disk) is found.}
\label{slope1}
\end{figure}

\begin{figure}[h!]
\centering
\includegraphics[width=3.0in]{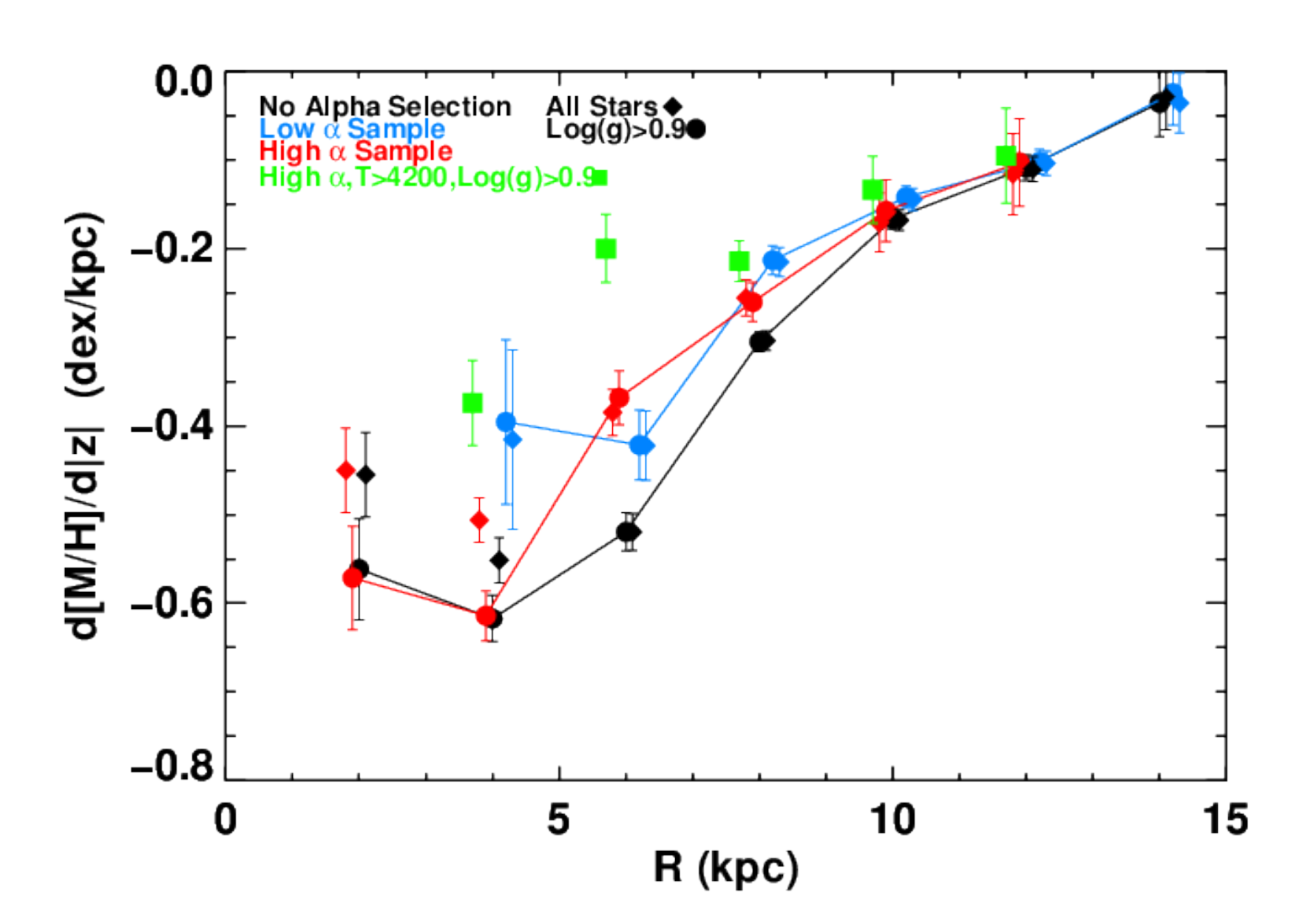}
\caption{The vertical gradient $\frac{\mathrm{d[M/H]}}{\textrm{d}|z|}$ for the APOGEE sample as a function of Galactocentric
radius. Results are shown for the full sample (black) as well as for the 
[$\alpha$/M] selected subsamples (blue for low-[$\alpha$/M], red and green for 
high-[$\alpha$/M]).  The measured slopes and their associated errors are 
slightly offset in radius to allow them to be more easily seen in the figure.  
The vertical gradients are steeper at smaller Galactocentric radii.
The vertical gradients are similar for both the low- and high-[$\alpha$/M] samples.}
\label{slope2}
\end{figure}

\begin{table*}[ht!]
  \caption{The observed vertical gradients and intercepts (at $z=0$) as a 
    function of $R$ for the different samples.}
  \vspace{1.5mm}
  \centering
  \begin{tabular}{c|cc|cc|cc|cc}
  \hline
  \hline
    Radial Range&\multicolumn{2}{|c|}{All Stars}&\multicolumn{2}{|c|}{Low $\alpha$ Stars}&\multicolumn{2}{|c|}{High $\alpha$ Stars}&\multicolumn{2}{|c}{High $\alpha$ Stars, T$_{\textrm{eff}}>4200~$K}\\
    (kpc)& ${\textrm{[M/H]}(z=0)}$ & d[M/H]/d\az & ${\textrm{[M/H]}}(z=0)$ & d[M/H]/d\az & ${\textrm{[M/H]}}(z=0)$ & d[M/H]/d\az & ${\textrm{[M/H]}}(z=0)$ & d[M/H]/d\az \\
    \hline
    $1<R<3$ & 0.39$\pm$0.04  & $-$0.56$\pm$0.06 & $-$ & $-$ & 0.39$\pm$0.04 & $-$0.57$\pm$0.06 & $-$ & $-$ \\
    $3<R<5$ & 0.31$\pm$0.02 & $-$0.62$\pm$0.03 & 0.30$\pm$0.11 & $-$0.40$\pm$0.09 & 0.28$\pm$0.02 & $-$0.61$\pm$0.03 & $-$0.04$\pm$0.20 & $-$0.37$\pm$0.05 \\
    $5<R<7$ & 0.27$\pm$0.01 & $-$0.52$\pm$0.02 & 0.27$\pm$0.03 & $-$0.42$\pm$0.04 & 0.02$\pm$0.03 & $-$0.37$\pm$0.03 & $-$0.19$\pm$0.06 & $-$0.22$\pm$0.04 \\
    $7<R<9$ & 0.04$\pm$0.01 & $-$0.31$\pm$0.01 & 0.03$\pm$0.02 & $-$0.21$\pm$0.02 & $-$0.09$\pm$0.03 & $-$0.26$\pm$0.02 & $-$0.17$\pm$0.05 & $-$0.22$\pm$0.02 \\
    $9<R<11$ & $-$0.15$\pm$0.01 & $-$0.17$\pm$0.01 & $-$0.15$\pm$0.01 & $-$0.14$\pm$0.01 & $-$0.25$\pm$0.03 & $-$0.16$\pm$0.04 & $-$0.28$\pm$0.03 & $-$0.13$\pm$0.04 \\
    $11<R<13$ & $-$0.29$\pm$0.01 & $-$0.11$\pm$0.02 & $-$0.29$\pm$0.01 & $-$0.10$\pm$0.02 & $-$0.32$\pm$0.04 & $-$0.10$\pm$0.05 & $-$0.33$\pm$0.03 & $-$0.10$\pm$0.05 \\
    $13<R<15$ & $-$0.43$\pm$0.02 & $-$0.04$\pm$0.04 & $-$0.43$\pm$0.01 & $-$0.03$\pm$0.04 & $-$ & $-$ & $-$ & $-$ \\
    \hline
  \end{tabular}
  \label{vertgrad}
\end{table*}

\subsubsection{Vertical Metallicity Gradient}

\label{sect:vertical}
\begin{figure*}[h!]
\centering
\includegraphics[width=5.0in]{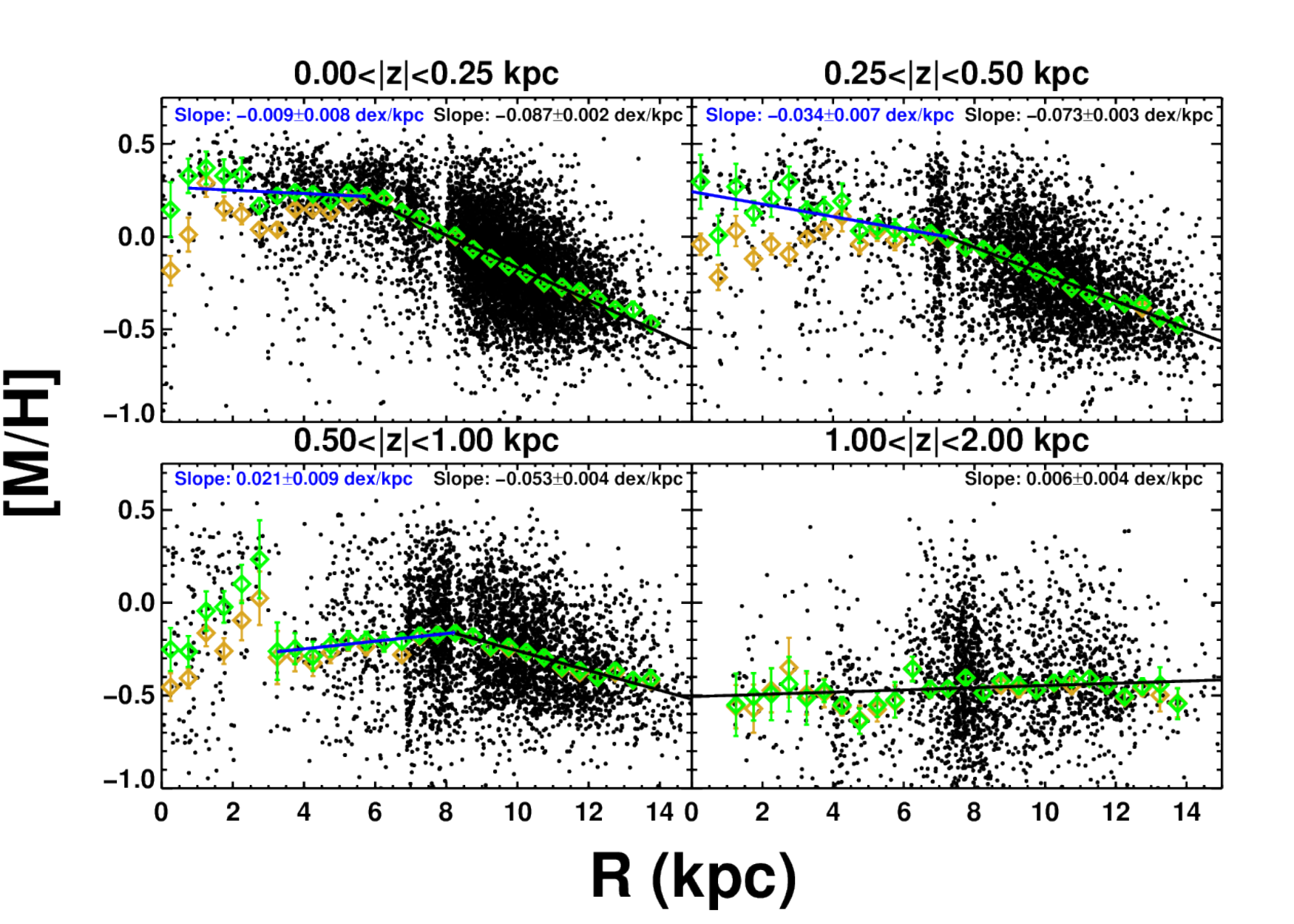}
\caption{The observed radial metallicity gradients over a range of \az\s for 
our sample of stars with $\log{textrm{g}}>0.9$. 
Green diamonds are the median metallicity for these stars,
while gold diamonds are the median metallicity for the entire sample 
(i.e., including stars with $\log{\textrm{g}} <0.9$).  The gradients 
presented in the text are fits to the green diamonds.  We find the gradient is steepest in the plane of the Galaxy, and the slope decreases with height.  In the inner Galaxy, the gradient becomes significantly less steep in the plane, with the transition at $R\sim6$ kpc.}
\label{radialgradall}
\end{figure*}

\begin{figure*}[h!]
\centering
\includegraphics[width=5.0in]{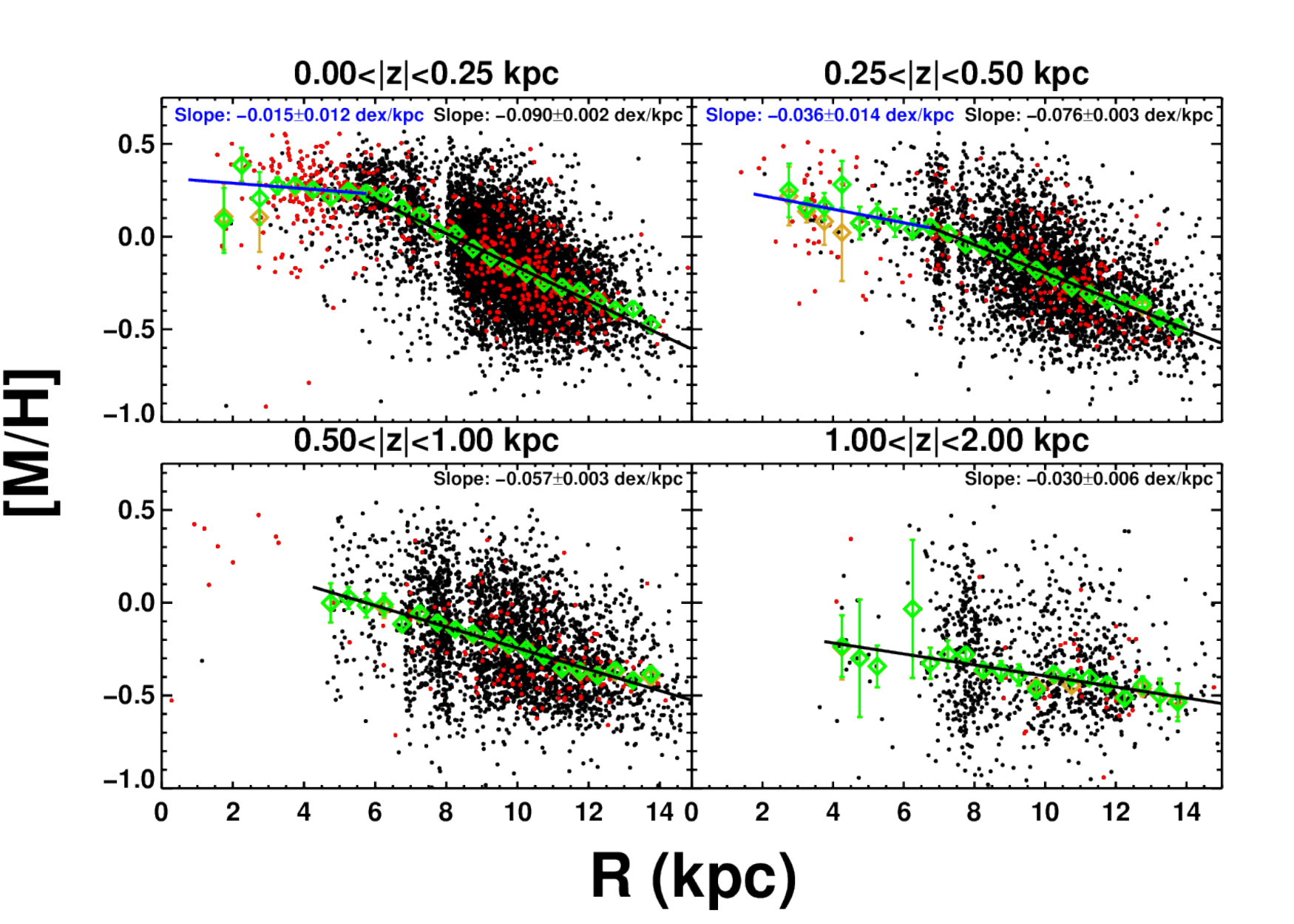}
\caption{The observed radial metallicity gradients over a range of \az\s for 
low-[$\alpha$/M] populations, after metallicities were corrected for the 
observed vertical gradient. Black dots provide the most
reliable and complete sample with $\log{\textrm{g}} >0.9$ and T$_{\textrm{eff}}>4200~$K; red dots are stars with T$_{\mathrm{eff}}<4200~$K.  The gold symbols are the same as in the previous figure.  The results are similar compared to that for the entire sample shown in Figure \ref{radialgradall}: the slope decreases with height about the plane, and the slope is shallower in the inner Galaxy with a transition at $R\sim6$ kpc.}
\label{radialgradlow}
\end{figure*}

\begin{figure*}[h!]
\centering
\includegraphics[width=5.0in]{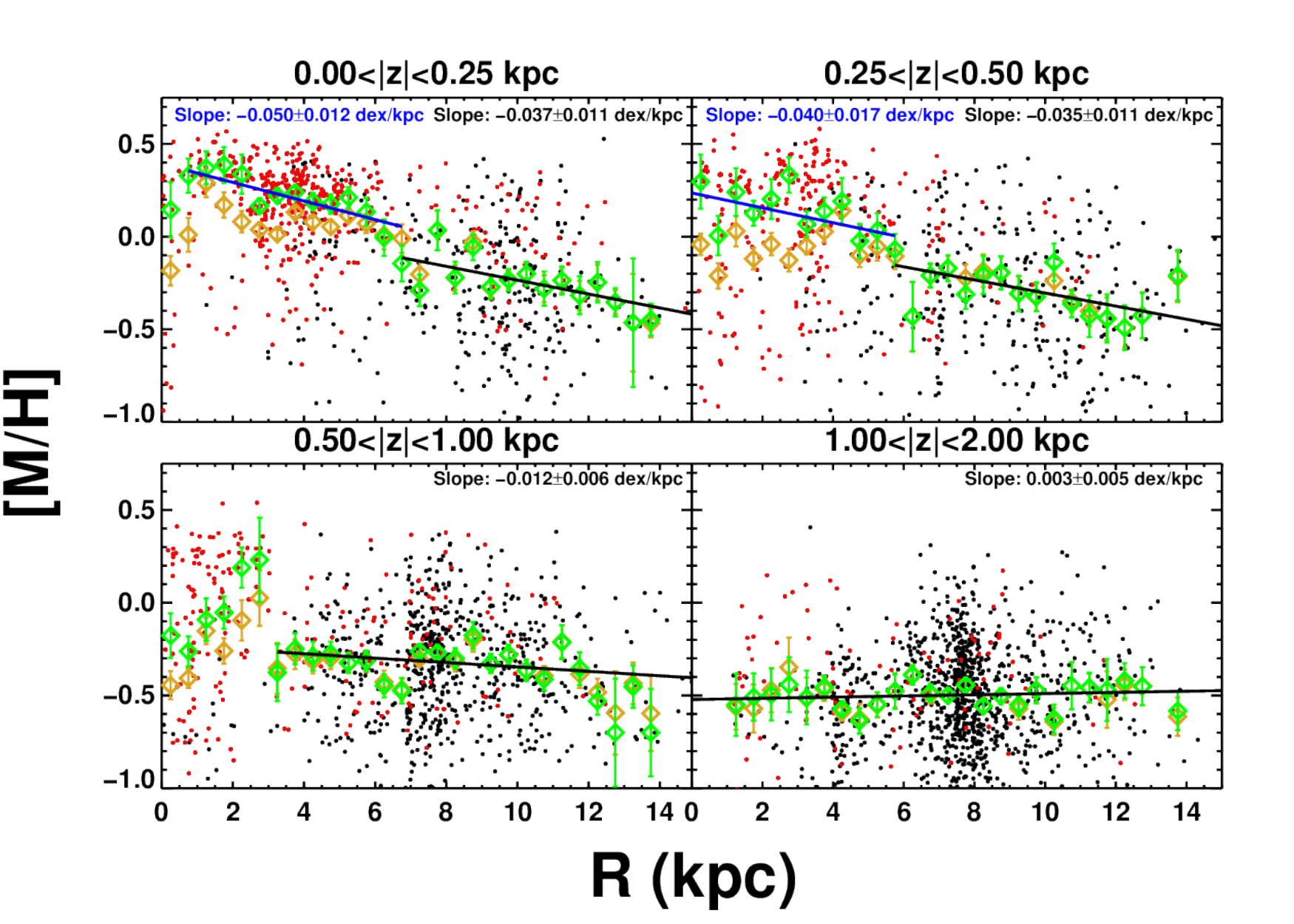}
\includegraphics[width=5.0in]{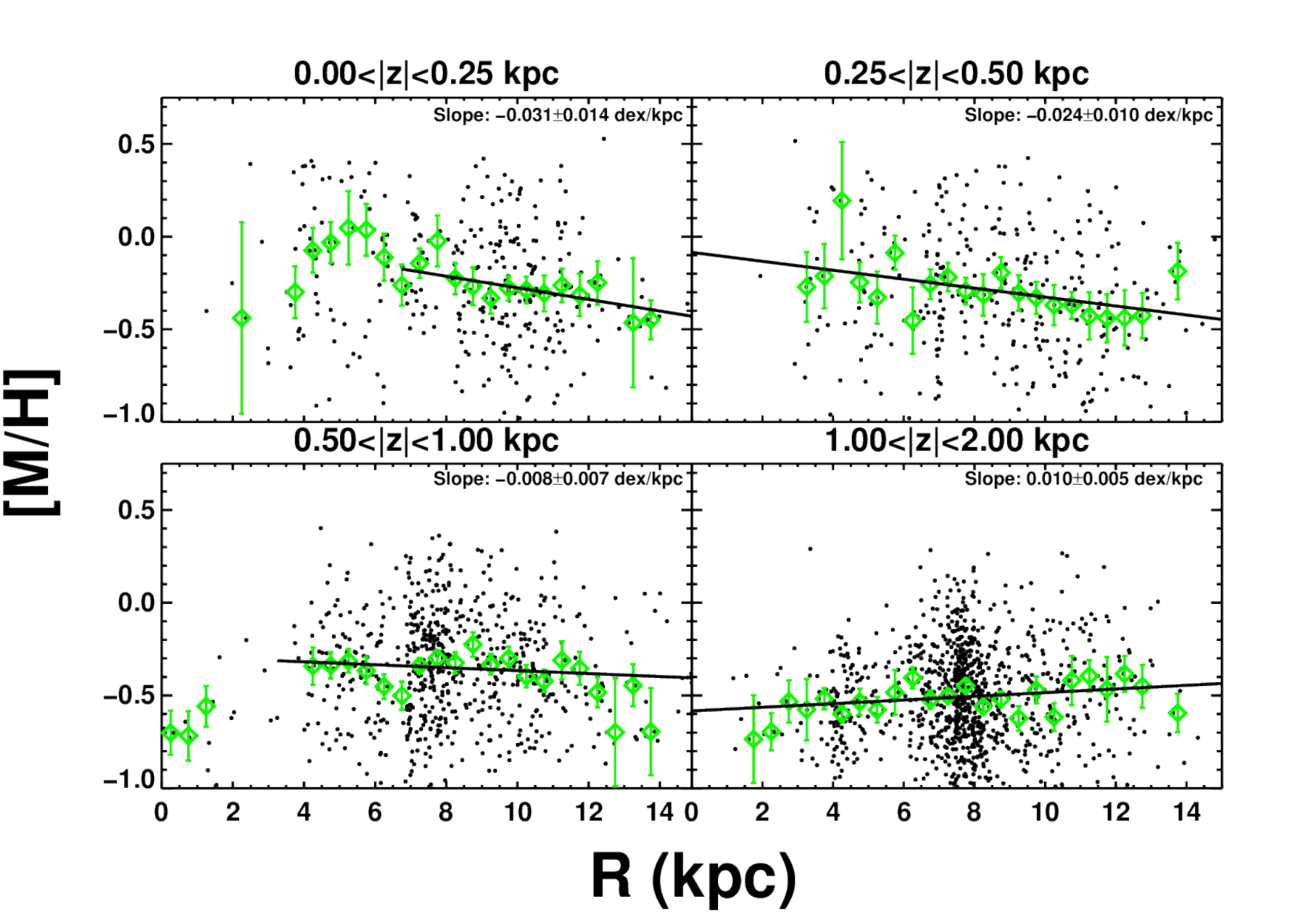}
\caption{\textit{Top}: Same as Figure \ref{radialgradlow}, but for high-[$\alpha$/M] subsamples. \textit{Bottom}: Similar to the top panel, but with stars cooler than 4200 K removed.  The slope of the gradient is shallower for the $\alpha$-enhanced samples compared to that of the low-[$\alpha$/M] sample, becoming flat for $|z|>0.5$ kpc.}
\label{radialgradhigh}
\end{figure*}

We measured vertical gradients from $0<|z|<2$ kpc in several radial bins
between 1 and 15 kpc; we used 2 kpc wide bins to ensure that there are
significant numbers of stars at all heights above the plane.  The vertical
gradient in each radial bin is measured by finding the median metallicity
[M/H] in 0.25 kpc intervals in $|z|$ and determining a linear fit to the
median values. Figure
~\ref{slope1} displays an example for the bin with $7<R<9$ kpc. The standard
error in the median is adopted as the uncertainty on each median point;
these uncertainties are quite small, given the large sample size.
A linear fit to the median metallicities appears to provide a good 
representation of the data.

The derived linear vertical gradients $\frac{\mathrm{d[M/H]}}{\textrm{d}|z|}$ as
a function of Galactocentric radius are shown in Figure \ref{slope2} for
the entire sample, as well as for the [$\alpha$/M] selected subsamples.

The vertical gradient is a smoothly changing function of Galactocentric
radius, becoming less steep at larger radii. The vertical gradient is
similar for all of the [$\alpha$/M] selected subsamples.  For the entire
sample, the vertical gradient is generally steeper than for either the
low or high-[$\alpha$/M] populations, because the low-[$\alpha$/M]
population has a higher mean metallicity and shorter vertical scale
height than the high-[$\alpha$/M] population \citep{Bovy2012a}.

Table \ref{vertgrad} presents the linear fits to the mean metallicities,
as a function of height above the plane, for several different zones of
Galactocentric radius for both the full sample and for the [$\alpha$/M]
selected subsamples. 
The vertical gradients are all well
characterized by a single linear fit.

\subsubsection{Radial Metallicity Gradient}
\label{sect:radial}

We measure the radial gradient in four different ranges of $|z|$:
$0<|z|<0.25$, $0.25<|z|<0.5$, $0.5<|z|<1$, and $1<|z|<2$ kpc between
$0<R<15$ kpc. Because of the significant vertical gradient, there can be
trends even within these bins (especially for the larger bins at larger
distance from the plane), so we shift all stars to a common \az\s
in each vertical bin by using the linear vertical abundance gradient fit
to correct the metallicities of individual stars to the bin center. For
each bin in \az, the radial gradient is then computed by determining
the median metallicity every 0.5 kpc between $0<R<15$ kpc. We use the
standard error in the median to characterize the uncertainty in median
metallicity in each radial bin. Figure \ref{radialgradall} presents the
radial gradients for entire APOGEE sample, while Figures \ref{radialgradlow}
and \ref{radialgradhigh} show the gradients for the low-
and high-[$\alpha$/M] populations.

For low-[$\alpha$/M] populations the radial gradients have a break
in the gradient for stars with $|z|<1$ kpc: the slope is flatter in the
inner Galaxy compared to the outer Galaxy. For stars
with $1<|z|<2$ kpc, however, a single component fit describes the
radial gradient, with a shallower slope.

The high-[$\alpha$/M] populations require a two-component fit for
$|z|<0.5$ kpc. However, it is possible that some of the high $\alpha$, 
high-metallicity stars with R$<6$ kpc may be contaminated by a bulge population, and 
that the [$\alpha$/M] abundance determination is not accurate for 
these stars, many of which have effective temperatures less than 4000 K.  

\begin{table*}[ht!]
  \caption{The observed radial gradients as a function of \az\s for the different subsamples.  The intercept of the fits are reported for $R=6$ kpc, which is the approximate radius at which the break occurs for most samples that have two components.}
  \vspace{1.5mm}
  \centering
  \begin{tabular}{c|ccc}
    \hline
    \hline
    \multicolumn{4}{c}{$0<|z|<0.25$ kpc} \\
    \multicolumn{1}{c}{Sample} & Radial Range (kpc) & ${\textrm{[M/H]}}(R=6 \textrm{kpc})$  & d[M/H]/d$R$ \\
    \hline
    All Stars& 0$<R<$5 & 0.21$\pm$0.06 & -0.009$\pm$0.008 \\
    All Stars& 5$<R<15$ & 0.18$\pm$0.02 & -0.087$\pm$0.002 \\
    Low $\alpha$ Stars& 0$<R<$5 & 0.23$\pm$0.10 & -0.015$\pm$0.012 \\
    Low $\alpha$ Stars& 5$<R<15$ & 0.20$\pm$0.02 & -0.090$\pm$0.002 \\
    High $\alpha$ Stars& 0$<R<$6 & 0.09$\pm$0.09 & -0.050$\pm$0.012 \\
    High $\alpha$ Stars& 6$<R<15$ & -0.09$\pm$0.13 & -0.037$\pm$0.011 \\
    High $\alpha$ Stars,T$_{\textrm{eff}}>4200~$K & 6$<R<15$ & -0.19$\pm$0.17 & -0.024$\pm$0.014 \\
    \hline
    \multicolumn{4}{c}{\rule{0pt}{2.75ex}$0.25<|z|<0.5$ kpc} \\
    \multicolumn{1}{c}{Sample} & Radial Range (kpc) & ${\textrm{[M/H]}}(R=6 \textrm{kpc})$ & d[M/H]/d$R$ \\
    \hline
    All Stars& 0$<R<$7 & 0.04$\pm$0.06 & -0.034$\pm$0.007 \\
    All Stars& 7$<R<15$ & 0.09$\pm$0.04 & -0.073$\pm$0.003 \\
    Low $\alpha$ Stars& 1$<R<$6 & 0.08$\pm$0.12 & -0.036$\pm$0.014 \\
    Low $\alpha$ Stars& 6$<R<15$ & 0.11$\pm$0.03 & -0.076$\pm$0.003 \\
    High $\alpha$ Stars& 0$<R<$5 & -0.01$\pm$0.12 & -0.040$\pm$0.017 \\
    High $\alpha$ Stars& 5$<R<15$ & -0.16$\pm$0.12 & -0.035$\pm$0.011 \\
    High $\alpha$ Stars,T$_{\textrm{eff}}>4200~$K & 0$<R<15$ & -0.23$\pm$0.10 & -0.024$\pm$0.010 \\
    \hline
    \multicolumn{4}{c}{\rule{0pt}{2.75ex}$0.5<|z|<1$ kpc} \\
    \multicolumn{1}{c}{Sample} & Radial Range (kpc) &  ${\textrm{[M/H]}}(R=6 \textrm{kpc})$ & d[M/H]/d$R$ \\
    \hline
    All Stars& 3$<R<$8 & -0.21$\pm$0.09 & 0.021$\pm$0.009 \\
    All Stars& 8$<R<15$ & -0.05$\pm$0.05 & -0.053$\pm$0.004 \\
    Low $\alpha$ Stars& 4$<R<15$ & -0.02$\pm$0.04 & -0.057$\pm$0.003 \\
    High $\alpha$ Stars& 3$<R<15$ & -0.30$\pm$0.06 & -0.012$\pm$0.006 \\
    High $\alpha$ Stars,T$_{\textrm{eff}}>4200~$K & 3$<R<15$ & -0.33$\pm$0.07 & -0.007$\pm$0.007 \\
    \hline
    \multicolumn{4}{c}{\rule{0pt}{2.75ex}$1<|z|<2$ kpc} \\
    \multicolumn{1}{c}{Sample} & Radial Range (kpc) & ${\textrm{[M/H]}}(R=6 \textrm{kpc})$ & d[M/H]/d$R$ \\
    \hline
    All Stars& 0$<R<15$ & -0.47$\pm$0.04 & 0.006$\pm$0.004 \\
    Low $\alpha$ Stars& 3$<R<15$ & -0.28$\pm$0.08 & -0.030$\pm$0.006 \\
    High $\alpha$ Stars& 0$<R<15$ & -0.50$\pm$0.05 & 0.003$\pm$0.005 \\
    High $\alpha$ Stars,T$_{\textrm{eff}}>4200~$K & 0$<R<15$ & -0.53$\pm$0.05 & 0.010$\pm$0.005 \\
    \hline
  \end{tabular}
  \label{radialgrad}
\end{table*}

Table \ref{radialgrad} lists the linear fits to the median metallicities
as a function of Galactocentric radius for different heights above
the plane for the full sample and for the [$\alpha$/M]
selected subsamples.
For the zones in which it appears that a single
linear fit cannot represent the data, we present 2-piece linear fits,
and specify the break radius in the table.  

\subsection{Robustness of the measured gradients}

The observed scatter in metallicity of $\pm0.2$ dex (at the one $\sigma$ level) observed in
Figures \ref{samples}, \ref{slope1}, \ref{radialgradall}, \ref{radialgradlow},
and \ref{radialgradhigh} at different locations in the Galaxy is real.
If the shape of the metallicity distribution function (MDF) changes as a
function of location, then the interpretation of gradients as determined
from the median metallicity might need to be revised; we will investigate
this issue in future work.

The small errors on the measured gradients result from the large
sample size observed by APOGEE, since we adopt the standard error
in the median for the uncertainty of the median metallicity in each bin.
We also characterized the gradients and their errors using a jack-knife
method \citep{Bootstrapref} which returned similar results (within one $\sigma$) and uncertainties as the standard error of the
median. We also performed linear fits to the raw metallicities (i.e., no binning) and
recovered similar gradients.

\section{Discussion}

\subsection{Comparison to Previous Literature: Mean Metallicities}

The APOGEE results extend previous measurements of mean metallicities to both
smaller and larger Galactocentric radii, and with coverage at different
heights above the plane over a range of Galactocentric radii. In particular,
the ability of APOGEE to probe dust-obscured populations allows our sample
to include stars in and near the Galactic plane at significant distances.

In regions covered by previous studies, we find similar mean
metallicities.  The mean metallicity of our low-[$\alpha$/M] sample
close to the plane of the Galaxy is similar to the metallicities found
by \citet{Luck2011a} in their sample of Cepheids at all Galactocentric
radii: they find mean metallicities of 0.25, 0.1, and $-0.3$
dex at $R$ of 6, 8, and 13 kpc respectively, in good agreement with
our measurements of 0.25, 0.05, and $-0.3$ dex for these radii. These
measurements are also consistent with the mean metallicities
observed by \citet{Frinchaboy2013} (using APOGEE open clusters,
which are distinct from the main survey targets considered here)
and \citet{2012AJ....144...95Y} in their surveys of Galactic open
clusters.

Away from the plane, \citet{Chen2011} measure a mean
metallicity of $-0.5$ dex for stars with $|z|>1$ kpc using SDSS/SEGUE
observations, which is the same value
that we measure for the APOGEE sample. For stars between $0.25<|z|<0.5$
kpc, we measure a slightly higher mean metallicity, $\langle\textrm{[M/H]}\rangle = -0.1$,
than \citet{Chen2011}, who measure a mean metallicity of $-0.2$.

\subsection{Vertical Metallicity Gradient}

The APOGEE data provide a first opportunity to characterize the vertical
metallicity gradient over a large range of Galactocentric radii.  As shown
in Figure \ref{slope2}, a vertical gradient exists at all Galactocentric
radii, but becomes shallower at larger R; the radial gradients become flatter at larger $|z|$. 

Our measurements are fairly consistent with previous results where
they overlap.  We measure a gradient of $-0.213\pm0.016$ dex
kpc$^{-1}$ for the low-[$\alpha$/M] sample at the solar circle.
Previous studies for similar populations of stars measure slightly
steeper slopes of $-0.295\pm0.005$ dex
kpc$^{-1}$ \citep{Chen2003}, $-0.23\pm0.04$ dex kpc$^{-1}$ \citep{2003BaltA..12..539B}, and
$-0.29\pm0.06$ dex kpc$^{-1}$ \citep{Marsakov2006}.  However, recent studies of open clusters have found little evidence for a vertical
gradient \citep{Carrera2011}, in contrast with earlier studies of open clusters \citep{Carraro1998,Carrera2011}. One potential explanation
for the discrepancy is earlier studies did not take into account the radial gradient when measuring the vertical gradient \citep{Carrera2011}. 
Our results, however, do take the radial gradient into account and are not compatible with measurements from \citet{Carrera2011}.  The open cluster
sample is small, however, and has very few objects more than 0.5 kpc above the plane, making this measurement heavily weighted by a few clusters. Additionally, the open
cluster sample is dominated by objects with ages $<1$ Gyr, so there are potentially age-metallicity effects between the two samples as well.

For the high-[$\alpha$/M] population
we measure a gradient of $-0.26\pm0.022$ dex kpc$^{-1}$ at the solar
circle, slightly higher than the slopes of $-0.22\pm0.07$ dex
kpc$^{-1}$ observed by \citet{Chen2011} and $-0.22\pm0.03$ dex
kpc$^{-1}$ by \citet{Ak2007} for their samples of thick-disk stars.
\citet {Soubiran2007} measure a slope of $-0.31\pm0.03 $ dex
kpc$^{-1}$ for their sample of red clump giants that span a range of
populations, in good agreement with our measurement of $-0.305\pm0.011$
dex kpc$^{-1}$ for the entire sample.

Interestingly, the vertical gradients appear to be similar for both low
and high-[$\alpha$/M] subsamples for $R>6$ kpc.  If the high-[$\alpha$/M]
stars represent an older sample, one might expect a different vertical
gradient for them if these gradients are the result of heating of a thin
disk population.

However, there is a significant difference in the vertical gradients at
$R<6$ kpc. Since the radial gradient in low-[$\alpha$/M] stars flattens
more than it is does for higher [$\alpha$/M] stars, the vertical gradient
becomes significantly steeper for high-[$\alpha$/M] stars in the inner
disk.

Several authors have found that older populations have larger
scale-heights, higher-[$\alpha$/M] ratios, and lower metallicity than
younger populations \citep{Lee2011,Bovy2012a,Schlesinger2012}.  
\citet{Minchev2013} measure
vertical gradients at the solar circle for their suite of simulations,
in which populations of the same age show no vertical gradient; they find
that observed vertical gradients in their simulations are due to mixing
of young, metal-rich populations that dominate the stellar density in
the plane with older, metal-poor populations that dominant the stellar
density above the disk.  If these results are true for the Milky Way,
the observed flattening of the vertical gradient with radius suggests that the
outer disk is more uniform in age, while the large vertical gradient in
the inner Galaxy imply a larger range of ages is present.

\subsection{Radial Metallicity Gradient}

The extended coverage of the APOGEE sample allows us to characterize the
radial gradient over a large range of radii at various heights above
the plane. 

\subsubsection{Inner regions}

While we find that there is a negative radial metallicity
gradient within a few kpc of the solar radius, this gradient appears to
flatten in the inner regions of the galaxy, at $R\lesssim 6$ kpc. This flattening
appears to be more significant in the low-[$\alpha$/M] sample than it
does in the high-[$\alpha$/M] sample. The change in slope from the
outer to inner regions is also stronger closer to the plane of the Galaxy 
than away from it.

Some degree of mixing of stars might be expected as a result of the
presence of a Galactic bar, and it is possible that our observed 
flattening of the gradient might be a sign of this effect.

\subsubsection{Intermediate radii}

At $R\gtrsim 6$ kpc, our measured radial gradient in the plane for low
[$\alpha$/M] stars of $-0.090\pm0.002$ dex kpc$^{-1}$ is in excellent
agreement with open cluster measurements from \citet{Frinchaboy2013},
who find a slope of $-0.09\pm0.03$ dex kpc$^{-1}$.  In that sample,
all but two of the clusters lie within 0.25 kpc of the plane.
\citet{Nordstrom2004} reported a radial gradient of $-0.099\pm0.011$
dex kpc$^{-1}$ for the intermediate age populations observed in the
GCS, again in good agreement with our measurement of $-0.090$ dex kpc$^{-1}$.

Considering the vertical dependence of radial gradients, we find
similar trends for $R\gtrsim 6$ kpc as \citet{Cheng2012b}: the gradients
become flatter as one moves away from the plane.  The gradient in the
plane, $-0.087\pm0.002$ dex kpc$^{-1}$, is steeper than that found
in SEGUE \citep{Cheng2012b}, who measure a slope of
$-0.066^{+0.030}_{-0.044}$ dex kpc$^{-1}$ in the plane, but this result
could partly arise from the fact that their sample was measured
with $0.15<|z|<0.25$, i.e., it did not go all the way down to b$=0^{\circ}$.
\citet{AllendePrieto2006} found no discernible gradient for their
sample of thick-disk stars between $1<|z|<3$ kpc, consistent with our
observations in our full sample for $|z|>1$ kpc. As one moves away
from the plane, the population becomes more metal-poor and
less variable with Galactocentric radius.

Our large sample, along with the ability to split the sample by
[$\alpha$/M], allows us to recognize that the radial gradient is
steeper for low-[$\alpha$/M] stars than for higher [$\alpha$/M]
stars at all distances from the plane; even in our $1<|z|<2$ kpc
bin, there are sufficient numbers of low-[$\alpha$/M] stars to demonstrate a 
negative radial gradient.

Although the measured radial gradients are much flatter for the
high-[$\alpha$/M] populations, our measurements demonstrate that this
population is chemically inhomogeneous in much the same way as the
low-[$\alpha$/M] populations close to the plane of the Galaxy, with the
radial gradient becoming flatter with height above the plane and
the vertical gradient becoming steeper in the inner Galaxy.

If one interprets the high-[$\alpha$/M] population as an older
population, our data suggest that the radial gradient is steeper
for younger stars than for older ones. However, while it is possible
that this could result from a metallicity gradient in the gas that
grows steeper with time, it could also result from the dilution of
a pre-existing gradient with time as might occur, for example, if
radial migration was important. There is disagreement in the
literature over this result. \citet{Stanghellini2010} observed
planetary nebula across a range of ages and find that older populations
have shallower metallicity gradients than younger populations.
However, in a similar study of a different set of planetary nebula,
\citet{Maciel2009a} find the opposite result; they observed the
gradient to flatten with time, with older populations having steeper
gradients. Recent models of chemical and galaxy evolution predict
that radial migration can wipe out gradients of older populations,
causing younger populations to have steeper observed gradients
(e.g., \citealt{Roskar2008,Loebman2011,Kubryk2013}).  Our result
of the steepening of the gradient with time are compatible with
these models and the results from \citet{Stanghellini2010}.

\subsubsection{Outer regions}

Previous studies (\citealt{Frinchaboy2013,2012AJ....144...95Y,Costa2004}) have
shown evidence for a transition to a flat radial gradient at large
Galactocentric radii in the plane, with the transition occurring between
10 and 13 kpc.  This flattening of the radial gradient is not observed
with Cepheids \citep{Luck2011a,Luck2011b,Lemasle2013a}.  There is some question
about whether the apparently flattening might arise if one does not
consider vertical gradients when the radial gradients are measured
\citep{Cheng2012b}.  We see no evidence for any flattening out to
$R\sim 14$ kpc, after correcting for vertical gradients, except perhaps
in the $0.5<|z|<1$ kpc bin.

Another possible explanation of the discrepancies found in previous
studies is that the gas densities are much lower in the outer Galaxy and
the chemical abundance is more sensitive to local events. The observations
at large Galactocentric radii are sensitive to the population(s) and
line(s) of sight being studied \citep{ Luck2011a}. With the uniform radial
coverage of APOGEE, we hope that the extension of the data set beyond
the first year data considered here will be useful for characterizing
the gradient at large radii.

When we consider the [$\alpha$/M] subsamples, we find that
the high-$\alpha$ population appears to truncate at $R\sim10-12$ kpc, as shown in Figures \ref{lowandhigh} and \ref{radialgradhigh}.
Previous studies of the $\alpha$-enhanced population have found that it 
has a shorter scale-length than the thin disk
\citep{Bensby2011,Cheng2012a}.  A detailed characterization of the
APOGEE selection function is still in progress, so we cannot
comment on the scale-length of the $\alpha$-enhanced populations, but
qualitatively, our observations are consistent with these previous studies.

\section{Conclusions} 

We present mean metallicity maps and chemical
abundance gradients observed in year 1 of SDSS-III/APOGEE for a
sample of nearly 20,000 red giants, and for subsamples that
split the sample by [$\alpha$/M].  Our primary results are:

\begin{itemize}
 \item Radial gradients exist in the plane, but become flatter or non-existent
above the plane.
 \item The radial gradient appears to flatten at $R<6$ kpc, more nearer the
disk than away from it, and more for low-[$\alpha$/M] stars.
 \item Radial gradients are steeper for low-[$\alpha$/M] stars than they
are for high-[$\alpha$/M] stars at all heights above the plane; a negative
gradient exists for the low-[$\alpha/M$] population even far from the plane,
while the gradient for high-[$\alpha/M$] stars disappears or even becomes
slightly positive far from the plane.
 \item Vertical metallicity gradients exist, but are flatter at larger 
Galactocentric radii.
 \item The vertical gradients are similar for low and high-[$\alpha$/M]
subsamples.
 \item There is real spread in the metallicities in all Galactic zones; 
the spread appears to be larger at smaller Galactocentric radii and
at larger distances from the mid-plane.
\end{itemize}

Detailed modeling will be required to understand how these results can
constrain models of disk formation and evolution, in particular, to understand
the roles of the evolution of the radial and vertical structure of the disk,
heating of disk populations, radial and vertical migration, and the importance
of external events (mergers or disturbances).

APOGEE is still an on-going project, and future observations will
expand the sample size and potentially expand the coverage
to larger radii and \az. The data reported in this paper represent only
a fraction
of the expected SDSS-III sample size. Additionally, improvements are being 
made in the chemical abundance determinations from ASPCAP. We expect
to be able to extend results to stars with T$_{\mathrm{eff}} < 3500$, better 
understand the $\alpha$-element abundances for stars with T$_{\mathrm{eff}}<4200$,
and to measure abundances for $\sim$ 15 separate elements with a goal
of 0.1 dex accuracy.

Funding for SDSS-III has been provided by the Alfred P. Sloan
Foundation, the Participating Institutions, the National Science
Foundation, and the U.S. Department of Energy Office of Science.
The SDSS-III web site is http://www.sdss3.org/.  SDSS-III is managed
by the Astrophysical Research Consortium for the Participating
Institutions of the SDSS-III Collaboration including the University
of Arizona, the Brazilian Participation Group, Brookhaven National
Laboratory, Carnegie Mellon University, University of Florida, the
French Participation Group, the German Participation Group, Harvard
University, the Instituto de Astrofisica de Canarias, the Michigan
State/Notre Dame/JINA Participation Group, Johns Hopkins University,
Lawrence Berkeley National Laboratory, Max Planck Institute for
Astrophysics, Max Planck Institute for Extraterrestrial Physics,
New Mexico State University, New York University, Ohio State
University, Pennsylvania State University, University of Portsmouth,
Princeton University, the Spanish Participation Group, University
of Tokyo, University of Utah, Vanderbilt University, University of
Virginia, University of Washington, and Yale University.

M.H, J.H. and S.M, acknowledge support for this research from
the National Science Foundation (AST-1109178).
Jo Bovy was supported by NASA through Hubble Fellowship grant HST-HF-
51285.01 from the Space Telescope Science Institute, which is
operated by the Association of Universities for Research in Astronomy,
Incorporated, under NASA contract NAS5-26555.
K.C. acknowledges partial support for this research from
the National Science Foundation (AST-0907873).

\bibliographystyle{apj}
\bibliography{ref}
\end{document}